\documentclass[12pt, draftclsnofoot, onecolumn]{IEEEtran}
\usepackage{cite}
\usepackage{amsmath,amssymb,amsfonts}
\usepackage{algorithm}
\usepackage{algorithmic}
\usepackage{graphicx}
\usepackage{textcomp}
\usepackage{xcolor}
\usepackage{multicol}
\usepackage{subcaption}
\usepackage{mathrsfs}
\usepackage{soul}
\usepackage{setspace}
\usepackage{listings}

\newcommand{\NR}[1]{{\color{blue} NR: #1}}
\begin{document}
\onecolumn
\title{Spectrum Shaping For Multiple Link Discovery in 6G THz Systems}
\author{\IEEEauthorblockN{Farzam Hejazi,}\textit{Department of Electrical and Computer Engineering}
\textit{University of Central Florida}\\
Orlando, USA, 
farzam.hejazi@ucf.edu\\
\and
\IEEEauthorblockN{ Katarina Vuckovic,}
\textit{Department of Electrical and Computer Engineering} 
\textit{University of Central Florida}\\
Orlando, USA,
kvuckovic@knights.ucf.edu\\
\and
\IEEEauthorblockN{ Nazanin Rahnavard,}{\textit{ Department of Electrical} \textit{and Computer Engineering,} 
\textit{University of Central Florida}\\
Orlando, USA,
nazanin@eecs.ucf.edu}
}

\maketitle
\begin{abstract}
     This paper presents a novel antenna configuration to measure directions of multiple signal sources at the receiver in a THz mobile network via a single channel measurement. Directional communication is an intrinsic attribute of THz wireless networks and the knowledge of direction should be harvested continuously to maintain link quality. Direction discovery can potentially impose an immense burden on the network that limits its communication capacity exceedingly. To utterly mitigate direction discovery overhead, we propose a novel technique called spectrum shaping capable of measuring direction, power, and relative distance of propagation paths via a single measurement. We demonstrate that the proposed technique is also able to measure the transmitter antenna orientation. We evaluate the performance of the proposed design in several scenarios and show that the introduced technique performs similar to a large array of antennas while attaining a much simpler hardware architecture. Results show that the spectrum shaping with only two antennas placed 0.5 mm, 5 mm, and 1 cm apart performs direction of arrival estimation similar to a much more complex uniform linear array equipped with 7, 60, and 120 antennas, respectively.    
\end{abstract}
\section{Introduction}

6G mobile networks promise to bring a new era of ultra high-speed communications that surpasses previous generations by several orders of magnitude in communication capacity \cite{dang2020should}. One of the core technologies behind such a spectacular revolution is massive Multi-Input-Multi-Output (MIMO) communication at mmWave and THz bands \cite{yang20196g}. Mobile networks that work at these frequency bands are bound to employ highly directional beams \cite{giordani2020toward}. As directional communication has gained importance in the new generation of communication systems, direction of arrival (DoA) estimation has obtained gravity as an enabler of directional communication \cite{chowdhury20206g}. To clarify this necessity, we should note that two devices that exploit directional antennas cannot communicate unless they ascertain the direction of the other device. Moreover, this knowledge of direction should be maintained during the communication period otherwise the link will be disrupted \cite{strinati20196g}.

The process of finding the best beam that maximizes the communication rate is called \emph{beam selection} \cite{kutty2015beamforming}. The problem of beam selection for mmWave and THz communication has been under extensive research  recently. The most common approach proposes a brute-force search on all beams \cite{ghasempour2017ieee}. Unfortunately brute-force search compels a huge overhead on the communication system since it should be harvested swiftly and continuously. As beams become narrower at higher frequencies, beam selection overhead escalates such that it drastically restricts the communication rate. Various techniques have been proposed to ease the burden of beam selection on communication systems. Ali et. al propose using legacy sub-6 GHz channel information for sparse recovery of mmWave channel and beam selection~\cite{ali2017millimeter}. Myers et. al utilize an efficient set of antenna weight vectors for fast beam alignment through a compressive sensing (CS) approach \cite{myers2019falp}. Swift-link \cite{myers2018swift} is a technique that incorporates randomized beam training along with a CS algorithm to develop beam selection technique robust to carrier frequency offset. Although these techniques can ease the overhead of beam selection, they still require numerous channel measurements. Many recent studies consider machine learning techniques to address the beam selection problem. Long et. al cast the beam selection problem as a multi-class classification problem and employ support vector machine (SVM) to achieve a statistical classification model that maximizes the sum rate \cite{long2018data}. Myers et. al propose an end-to-end deep learning technique to design
a structured CS matrix based on the underlying channel distribution, leveraging both sparsity and the particular spatial structure of propagation paths that appears in a communication channel~\cite{myers2020deep}. Alrabeiah et al. train a deep network to learn the mapping between sub-6 GHz channel state information (CSI) and optimal mmWave beam \cite{alrabeiah2020deep}. Recently, several works consider vision, LIDAR, Radar, and other means of situational awareness to integrate with communication data for mmWave channel estimation and beam selection \cite{xu20203d,charan2020vision, alrabeiah2020millimeter,klautau2019lidar}. 
A common limitation among available literature incorporating deep learning for beam selection is the blindness to unseen data and lack of adaptability to time-varying dynamic environments.

In the most recent breakthrough
, Ghasempour et. al. introduce the idea of THz rainbow for single-shot link discovery \cite{ghasempour2020single}. In this work, authors name the process of finding the DoA of signal from the transmitter (TX) via only a single measurement, single-shot link discovery.  The authors employ leaky wave antennas at both TX and receiver (RX). When excited by a broadband source (i.e. a pico second pulse with a flat spectrum between [0.1THz,1THz]), the TX antenna propagates a different and unique frequency within the spectrum at each angle, thus forming a THz rainbow. The RX employs THz time-domain spectroscopy (THz-TDS) to measure the spectrum of the received signal and consequently estimates its DoA. THz-TDS is a technique to measure the THz electric field that can be used to measure the received spectrum in the range [0.1THz,10THz] and currently is implementable on chip \cite{globisch2015absolute}. Other than DoA, the THz rainbow technique is able to measure the angle of departure (AoD) of the signal from the TX, yet in a limited range. Although measuring DoA and AoD in a single shot is a groundbreaking achievement, the THz rainbow technique suffers from following inefficiencies: {

\begin{itemize}
    \item very limited DoA and AoD observability range ($[10^o,80^o]$ for DoA, and $DoA + [-40^o,20^o]$ for AoD),
    \item incapablity in estimating multiple DoAs, thus inapplicable in harsh multipath environments,
    \item limited applicability to only dry indoor environments 
    \item spectral inefficiency as a result of using only a very tiny range of available spectrum for DoA estimation. 
\end{itemize}

In our work, we propose a new design to address the inefficiencies of THz rainbow, which we refer to as \emph{Spectrum Shaping (SSH)}. We adopt a well-known array design with antenna spacing that typically makes use of the time difference of arrival (TDoA) of signals between two antennas to estimate DoA in LoS communication \cite{alarifi2016ultra,heydariaan2020anguloc }. However, since we reduce the antenna spacing to less than a centimeter, measuring TDoA requires several Tera-samples per seconds which is not accessible. On the other hand, instead of measuring TDoA, by carefully adding delay lines in the design we try to shape the received spectrum in such a way that enables us to estimate multiple concurrent DoAs leveraging the spectrum of the spectrum of the received signal. Moreover, we devise a novel antenna placement at the receiver and the TX to estimate AoD and DoA using only a single measurement. To the best of our authors knowledge, this is the first time that such an antenna spacing design is introduced for AoD and DoA estimation in a single-shot for THz band applicable in the presence of strong multi-path. In spectrum shaping, we incorporate the same broadband signal source at the TX and TDS-THz at RX as the THz rainbow. The main characteristics of our novel design can be encapsulated as  

\begin{itemize}
    \item capability of measuring DoA and AoD in a wide range ($[0^o,180^o]$),
    \item capability of measuring multiple incoming signals with multiple DoAs in a single shot,
    \item performing similar to a large array while benefiting from a much simpler architecture,
    \item utilizing the whole available spectrum for DoA estimaton, 
    \item no requirement for synchronization between RX and TX,
    \item applicability in humid indoor and outdoor environments, and
    \item resiliency to attenuation due to atmospheric gases and presence of intense water vapor.
\end{itemize}

The rest of the paper is organised as follows. In Section \ref{sysmod}, we present the system model and our novel RX design. We demonstrate the capability of the proposed design in measuring multiple DoAs in a single shot. We propose a new TX design and show the capability of the system in measuring DoA and AoD in a wide angular range in Section \ref{DoAAoD}. We derive the Cramer Rao Lower Band of error for DoA and AoD estimation in Section \ref{CRBSec}. In Section \ref{sim}, we demonstrate the performance of spectrum shaping by several simulations and show that it performs similar to a large antenna array. Finally, we conclude the paper in Section \ref{conc}.  }      

\section{System model}
\label{sysmod}

\begin{figure}[htb]
    \centering
    \includegraphics[width=4in,height=2in]{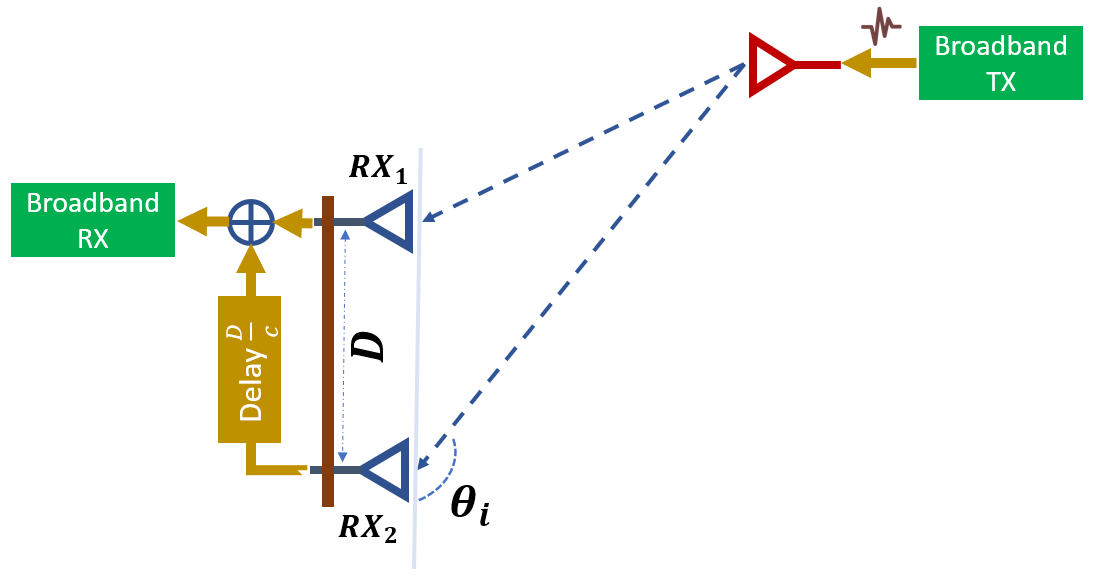}
    \caption{Block diagram of the spetrum shaping for detecting incoming DoAs.} 
    \label{BDones}
\end{figure}

As illustrated in Fig. \ref{BDones}, we consider a link where RX is equipped with an antenna pair. The gap between the two antennas is denoted by $D$. The transmitter (TX) is excited by a broadband source generating a single pulse of broadband emission whose spectral coverage is broad enough to cover the entire relevant band (0.1 to 1 THz). The first ($RX_1$) and the second ($RX_2$) antennas receive the signal emitted by the TX. After passing through a delay line with length $D$, the received signal at $RX_2$ is superimposed with the signal received at $RX_1$ and fed into a THz-TDS receiver.

\subsection{Line of Sight (LoS) DoA Estimation}
\label{secA}
In this section, we assume there is only one LoS path between the TX and the RX. We denote the DoA of the signal to the RX by $\theta_{i}$. Given the far field assumption, the signal emitted by the TX is received at the two antennas of the RX with a time shift $\delta t_i$ given by 

\begin{equation}
    \delta t_i = - \frac{D}{c}\cos(\theta_i) \,,
\end{equation}

where $c$ is the speed of light. The broadband RX then measures the spectrum of the superimposition of the signals received at $RX_1$ and $RX_2$ (with a delay equal $\frac{D}{c}$) using THz-TDS technique. Subsequently, the output spectrum of the broadband receiver can be expressed as 
\begin{align}
\label{spes}
    &\boldsymbol{E}_r(f,\theta_i) = |\mathcal{F}\{ r(t) + r(t-\frac{D}{c}-\delta t_i)  \}|^2=|R(f)+R(f) e^{-j2\pi f (\delta t_i+\frac{D}{c})}|^2 \\ \nonumber &=  |2R(f)|^2|cos(\pi f \frac{D}{c}(1-\cos(\theta_i)))|^2\\ \nonumber
    &= |2R(f)|^2|cos(\pi f \frac{2D}{c}\sin^2(\frac{\theta_i}{2}))|^2 \,,
\end{align}
where $\boldsymbol{E}_r(f,\theta_i)$ is the spectrum of the superimposition of the two received signals, $\mathcal{F\{.\}}$ is the Fourier transform operator, $r(t)$ is the received signal at the $RX_1$, and $R(f)$ is the Fourier transform of the $r(t)$. We assume only one LoS path between the RX and the TX; hence, $r(t) = a(t)*s(t)$, where $a(t)$ models THz channel (which is typically a frequency selective channel at the interested band), $s(t)$ is the transmitted signal, and $*$ is the convolution operator. Considering $s(t)$ has a flat spectrum over the relevant band, \eqref{spes} can be expressed as
\begin{equation}
    \boldsymbol{E}_r(f,\theta_i) =  C|a(f)|^2 + C|a(f)|^2 \cos(2\pi f \frac{2D}{c}\sin^2(\frac{\theta_i}{2}))\,.
    \label{arres1}
\end{equation}
where $C$ is a constant and $a(f)$ is the channel frequency response. Applying Fourier transform over $\boldsymbol{E}_r(f,\theta_i)$, defining $\mathbf{a}(\zeta)=\mathcal{F}\{|a(f)|^2\}$ we have 
\begin{equation}
    \mathcal{F}\left\{\boldsymbol{E}_r(f,\theta_i)\right\} =  C  \left(\mathbf{a}(\zeta) + \frac{1}{2}\mathbf{a} \left(\zeta-\frac{2D}{c}\sin^2(\frac{\theta_i}{2})\right)+ \frac{1}{2}\mathbf{a} \left(\zeta+\frac{2D}{c}\sin^2(\frac{\theta_i}{2})\right)\right)\,.
    \label{arres2}
\end{equation}

{$\mathbf{a}(\zeta)$ is the frequency response of the THz channel frequency response. Fig. \ref{Drych} and \ref{HumidCH} depict $\mathbf{a}(\zeta)$ for a dry (zero water vapor density) and a humid (water vapor density equals $10\frac{g}{m^3}$) environments, respectively. Although, channels are frequency selective, $\mathbf{a}(\zeta)$ shows a very strong and distinctive global pick at zero and at other frequencies is roughly zero. In Section \ref{sim}, we will show by simulation that THz channel can potentially result in a reduction of angular resolution of the technique in humid environments which can be compensated by increasing $D$. Since $\sin^2(\frac{\theta_i}{2})$ is positive,  $\sin^2(\frac{\theta_i}{2})$ can be simply estimated via finding the only positive element of the spectrum of $\boldsymbol{E}_r(f,\theta_i)$. Then if $\theta_i \in [0,\pi]$, $\theta_i$ can be uniquely estimated. Moreover, since we exclusively make use of the amplitude of the Fourier transform of the received signal (and not its phase), we eliminate the requirement of synchronization between the TX and the RX \cite{ghasempour2020single}. It should be mentioned that, without the delay line after $RX_2$, the $\boldsymbol{E}_r(f,\theta_i)$ turns out to be  
\begin{equation}
    \boldsymbol{E}_r(f,\theta_i) =  C |a(f)|^2 (1+\cos(2\pi f \frac{D}{c}\cos(\theta_i)))\,.
    \label{arres2}
\end{equation}
In this case, $\boldsymbol{E}_r(f,\theta_i) = \boldsymbol{E}_r(f,-\theta_i) = \boldsymbol{E}_r(f,\pi - \theta_i)$. Therefore, DoA is uniquely observable \emph{only} if $\theta_i \in [0,\frac{\pi}{2}]$. Adding the delay line, we \emph{double} the DoA observability range to $[0,{\pi}]$. Regarding \eqref{spes}, we observe that the proposed design ends up in multiplication of $|\cos(\pi f \frac{2D}{c}\sin^2(\frac{\theta_i}{2}))|^2$ to the received spectrum, which provide us with enough information to estimate DoA in the whole range from $0^o$ to $180^o$. Hence, we call the multiplied term \emph{spectrum shaper} and the proposed technique \emph{spectrum shaping}.   

\begin{figure}[h!]
    \centering
    \includegraphics[width=4in,height=3.4in]{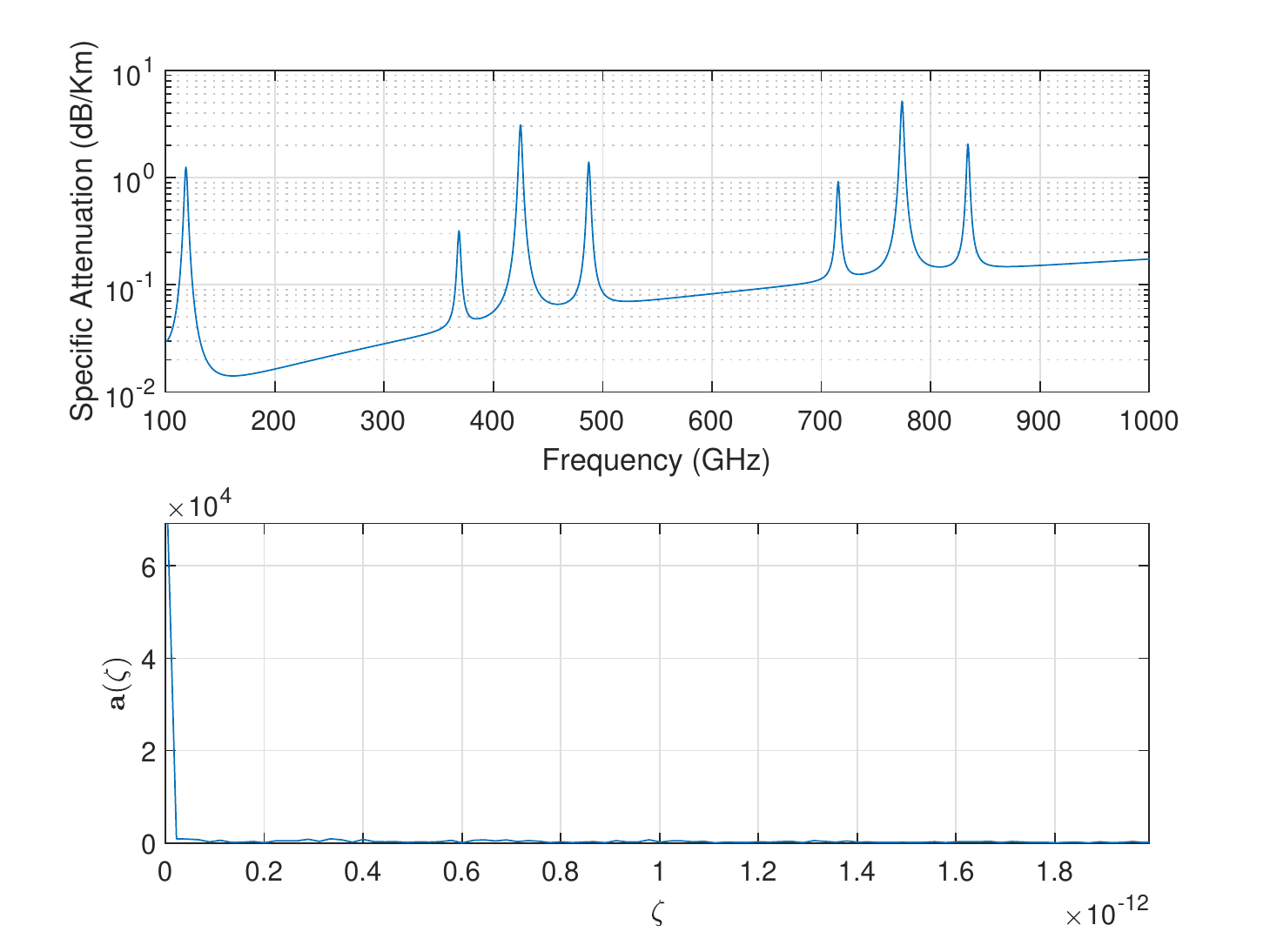}
    \caption{Specific attenuation due to atmospheric gases (above figure) and frequency response of the frequency response of the channel in a dry environment (typically an indoor environment). Although THz channel is naturally a frequency selective channel, the rate of change of the channel based on frequency is very slow, thus the frequency response of the frequency response of it contains a sole distinctive peak at zero.} 
    \label{Drych}
\end{figure}

\begin{figure}[htb]
    \centering
    \includegraphics[width=4in,height=3.4in]{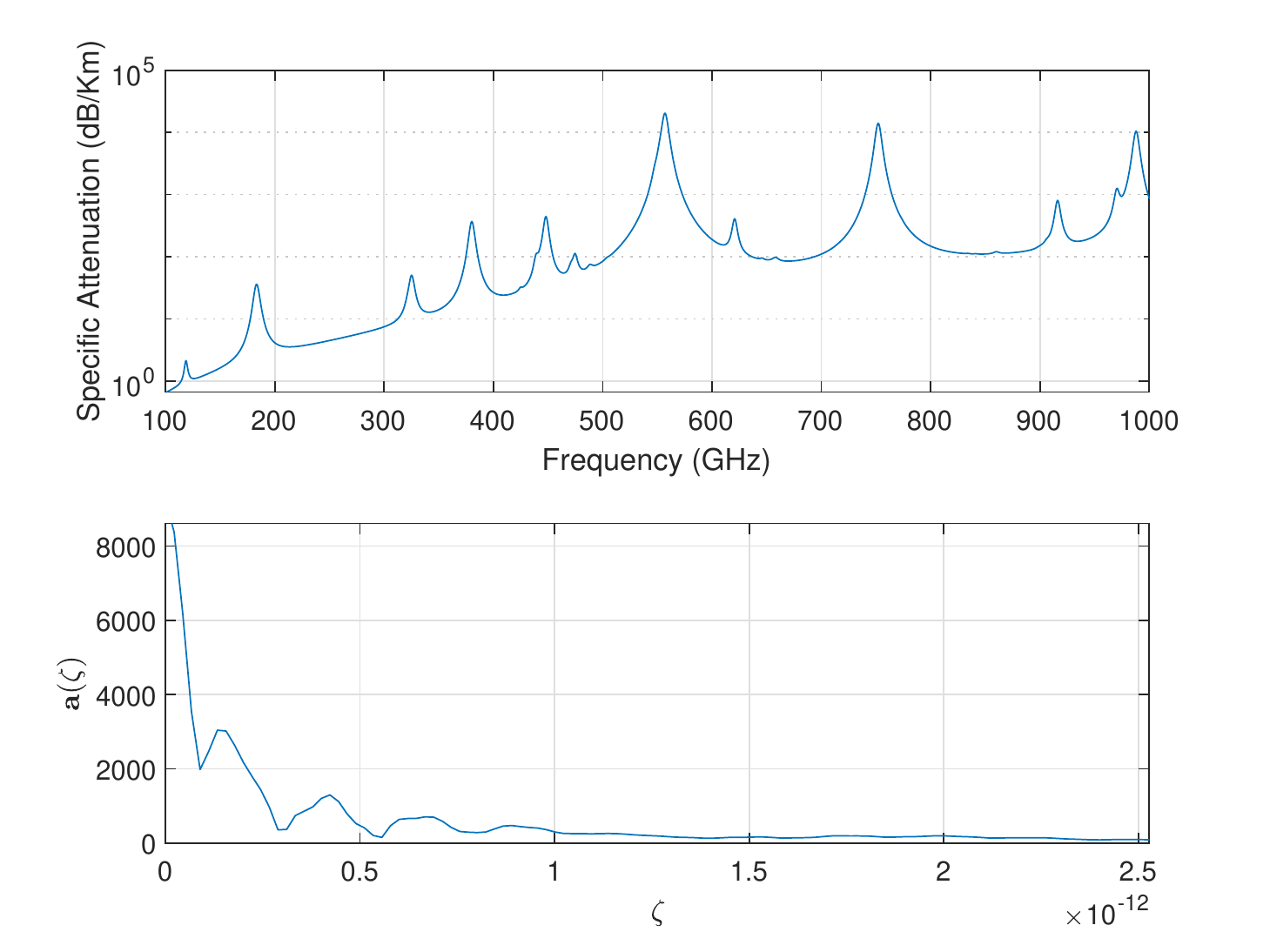}
    \caption{ Specific attenuation due to atmospheric gases (above figure) and frequency response of the frequency response of the channel in a humid environment (path length is 1000m). The rate of change of the channel versus frequency is more than a dry environment,  however, it is relatively slow such that   frequency response of it's frequency response contains a sole distinctive peak at zero.} 
    \label{HumidCH}
\end{figure}

}

\subsection{ Multiple Non-LoS (NLoS) DoA Estimation}
\label{Mult}

In this section, we assume that there are multiple paths between the TX and RX (or similarly there are multiple TXs in the environment). We intend to detect powers and angles of all incoming paths to the RX. Suppose there are $m$ paths between the TX and the RX and $a_k(t), \theta_k, T_k; k \in \{1,\dots,m\}$ denote the THz channel through path $k$, DoA of path $k$, and the time of flight (ToF) of the signal through path $k$, respectively (Fig. \ref{BDones}). 
 The received signal at the RX antenna pair can be expressed as 
\begin{figure}[htb]
    \centering
    \includegraphics[width=4.3in,height=2.3in]{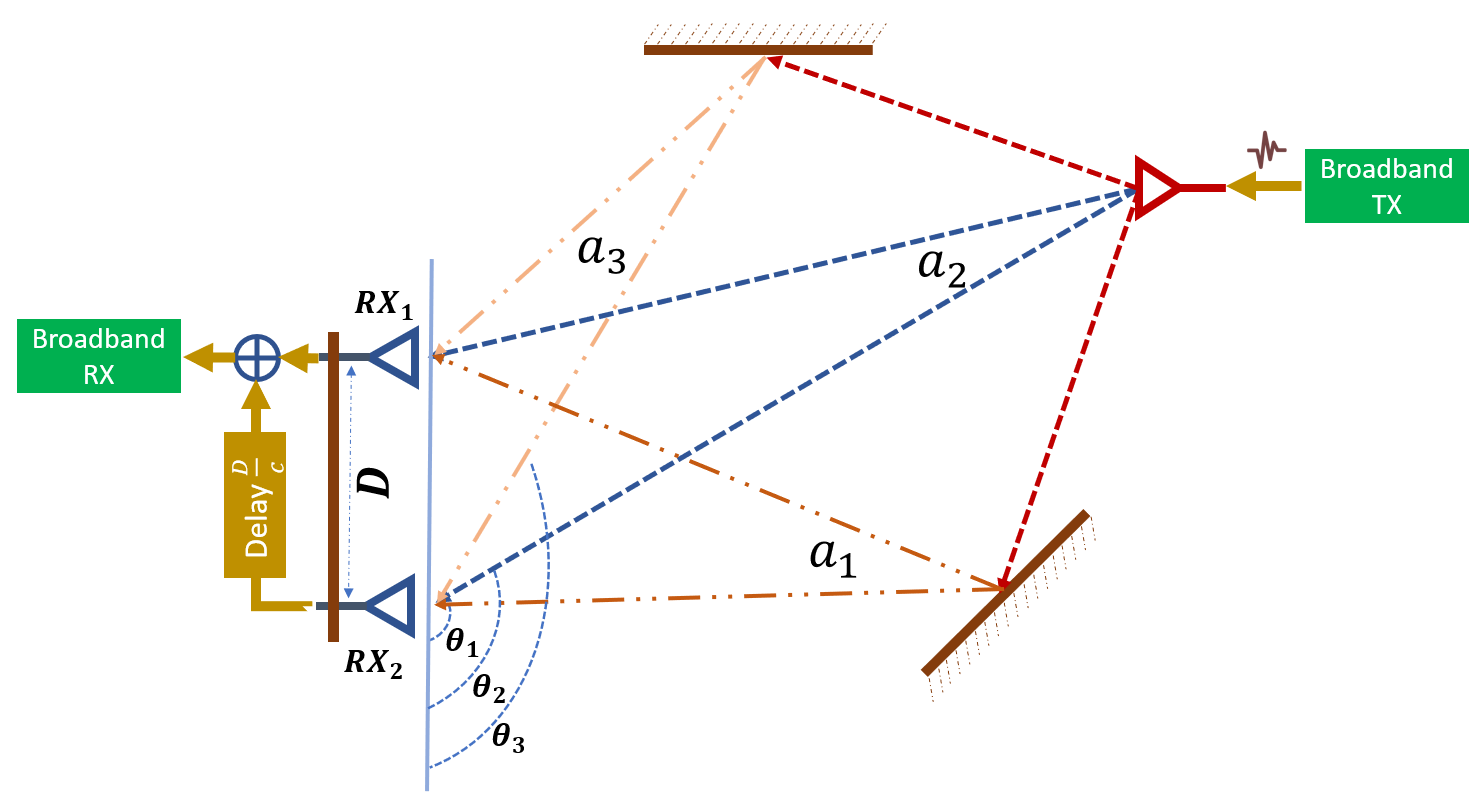}
    \caption{ spectrum shaping can be employed for detecting multiple incoming DoAs. There are 3 paths between the TX and The RX. $\theta_1,\theta_2,\theta_3$ and $a_1,a_2,a_3$  are DoAs the and path loss of the three paths, respectively.} 
    \label{BDmult}
\end{figure}
\begin{align}
    &r_1(t) = \sum_{k=1}^{m} a_k(t) * s(t-T_k) \\  r_2(t) = &\sum_{k=1}^{m} a_k(t)* s(t-T_k+\frac{D}{c}\cos(\theta_k)) \,,
\end{align}

where $r_1(t)$ and $r_2(t)$ are the received signals at $RX_1$ and $RX_2$, respectively. Passing $r_2(t)$ through the delay line in Fig. \ref{BDones}, the spectrum of the superimposition of the two signal turns out to be 

\begin{align}
    &\boldsymbol{E}_r(f, \theta_1,\dots,\theta_m) = |\mathcal{F}\{ r_1(t) + r_2(t-\frac{D}{c})\}|^2 \nonumber \\ &=\left|\sum_{k=1}^{m} a_k(f) S(f) e^{-j2\pi f T_k} + \sum_{k=1}^{m} a_k(f) S(f) e^{-j2\pi f T_k} e^{-j2\pi f \frac{D}{c}(1-\cos(\theta_k))}\right|^2 \nonumber \\
    &=|2S(f)|^2 \left|\sum_{k=1}^{m} a_k(f) e^{-j2\pi f (T_K+\frac{D}{c}\sin^2(\frac{\theta_k}{2}))} \cos{(\pi f \frac{2D}{c} \sin^2(\frac{\theta_k}{2}))}\right|^2 \nonumber \\
    &= |2S(f)|^2 (\sum_{k=1}^{m} a^2_k(f)\cos^2(\pi f \frac{2D}{c} \sin^2(\frac{\theta_k}{2})) 
    + \nonumber \\ & \sum_{k=1}^{m}\sum_{p=k+1}^{m} 2a_k(f) a_p(f) \cos{(\pi f \frac{2D}{c} \sin^2(\frac{\theta_k}{2}))} \cos{(\pi f \frac{2D}{c} \sin^2(\frac{\theta_p}{2}))}
    \cos{(2\pi f (T^{\prime}_k - T^{\prime}_p))} )
    \,.
    \label{detailedspecs}
\end{align}

where $T^{\prime}_k = T_k+\frac{D}{c}\sin^2(\frac{\theta_k}{2}), k \in {1,\dots,m} $. Given $s(t)$ has a flat frequency response over the relevant band, $\boldsymbol{E}_r(f, \theta_1,\dots,\theta_m)$ is proportional to 

\begin{align}
    &\boldsymbol{E}_r(f, \theta_1,\dots,\theta_m) \propto \nonumber \\ 
    &(\sum_{k=1}^{m} \frac{a^2_k(f)}{2} + \sum_{k=1}^{m} \frac{a^2_k(f)}{2}\cos(2 \pi f \frac{2D}{c} \sin^2(\frac{\theta_k}{2})) +\nonumber \\  &\sum_{k=1}^{m}\sum_{p=k+1}^{m} a_k(f) a_p(f) \cos{(\pi f \frac{2D}{c} \sin^2(\frac{\theta_k}{2}))} \cos{(\pi f \frac{2D}{c} \sin^2(\frac{\theta_p}{2}))}
    \cos(2\pi f (T^{\prime}_k - T^{\prime}_p)))
    \,.
    \label{spectp}
\end{align}

Considering \eqref{spectp}, $\boldsymbol{E}_r(f, \theta_1,\dots,\theta_m)$ is a summation of multiple cosine functions modulated with different frequencies\footnote{The third term in \eqref{spectp} consists of multiplications of three cosine terms. From basic trigonometry we know that any arbitrary multiples of cosines equals sum of cosines. }, hence, the spectrum of  $\boldsymbol{E}_r(f, \theta_1,\dots,\theta_m)$  exhibit multiple spikes corresponding to each cosine function.

$\boldsymbol{Lemma\: 1.}$ Assuming $|T^{\prime}_k - T^{\prime}_p| \gg \frac{2D}{c}; p \neq k ;p,k \in \{1,\dots,m\} , $ the third term in \eqref{spectp} can be filtered out by applying a low-pass filter with cut-off frequency $\frac{2D}{c}$ on $\boldsymbol{E}_r(f, \theta_1,\dots,\theta_m)$.

\textbf{Proof.} The frequency of all the elements in the first and the second terms of \eqref{spectp} is less than $\frac{2D}{c}$, while all the elements in the third term are modulated by $\cos(2\pi f (T^{\prime}_k - T^{\prime}_p))$ harmonics. Thus, if the condition in Lemma 1 holds, the third term can be removed using a low-pass filter with cut-off frequency $\frac{2D}{c}$. $\square$

The condition in Lemma 1 expresses that the third term of \eqref{spectp} is removable by filtering if the difference between ToF of all paths to the RX are far larger than $\frac{2D}{c}$; or equivalently the difference between paths' length of all paths to the RX are far larger than $2D$. In Section \ref{sim}, we will discuss that $D$ is typically in the sub centimeter range (i.e. 1 mm to 1 cm). Thus the condition of Lemma 1 will hold in most practical use cases. After passing $\boldsymbol{E}_r(f, \theta_1,\dots,\theta_m)$ through a low-pass filter with cut-off frequency $\frac{2D}{c}$, referring to \eqref{arres1}, the output $\boldsymbol{\hat{E}}_r(f, \theta_1,\dots,\theta_m)$ can be expressed as 

\begin{align}
   \boldsymbol{\hat{E}}_r(f, \theta_1,\dots,\theta_m) = C \left (
    \sum_{k=1}^{m} {a^2_k(f)} + \sum_{k=1}^{m} {a^2_k(f)}\cos(2 \pi f \frac{2D}{c} \sin^2(\frac{\theta_k}{2})) \right ) =  \sum_{k=1}^{m} \boldsymbol{E}^{(k)}_r(f,\theta_k)
    \,,
    \label{spectp1}
\end{align}
where $\boldsymbol{E}^{(k)}_r(f,\theta_k)$ is the spectrum of the received signal if only one LoS path arrived at the antenna pair from path $k$. According to \eqref{spectp1}, the output spectrum of THz-TDS  after lowpass filtering will result in a linear weighted summation of $\boldsymbol{E}^{(k)}_r(f,\theta_k)$'s. In other words, the proposed design is a linear system facing multiple incoming signals.    
Since $\boldsymbol{\hat{E}}_r(f, \theta_1,\dots,\theta_m)$ composes of linear summation of harmonic signals, by using harmonic decomposition techniques  such as Fourier transform, MUSIC \cite{schmidt1986multiple}, or Pisarenco harmonic decomposition \cite{pisarenko1973retrieval} we can separate and detect different harmonics of $\boldsymbol{\hat{E}}_r(f, \theta_1,\dots,\theta_m)$ and consequently we can determine the angles and the powers of the incoming paths.

\section{AoD Estimation}
\label{DoAAoD}
\begin{figure}[htb]
    \centering
    \includegraphics[width=4in,height=2in]{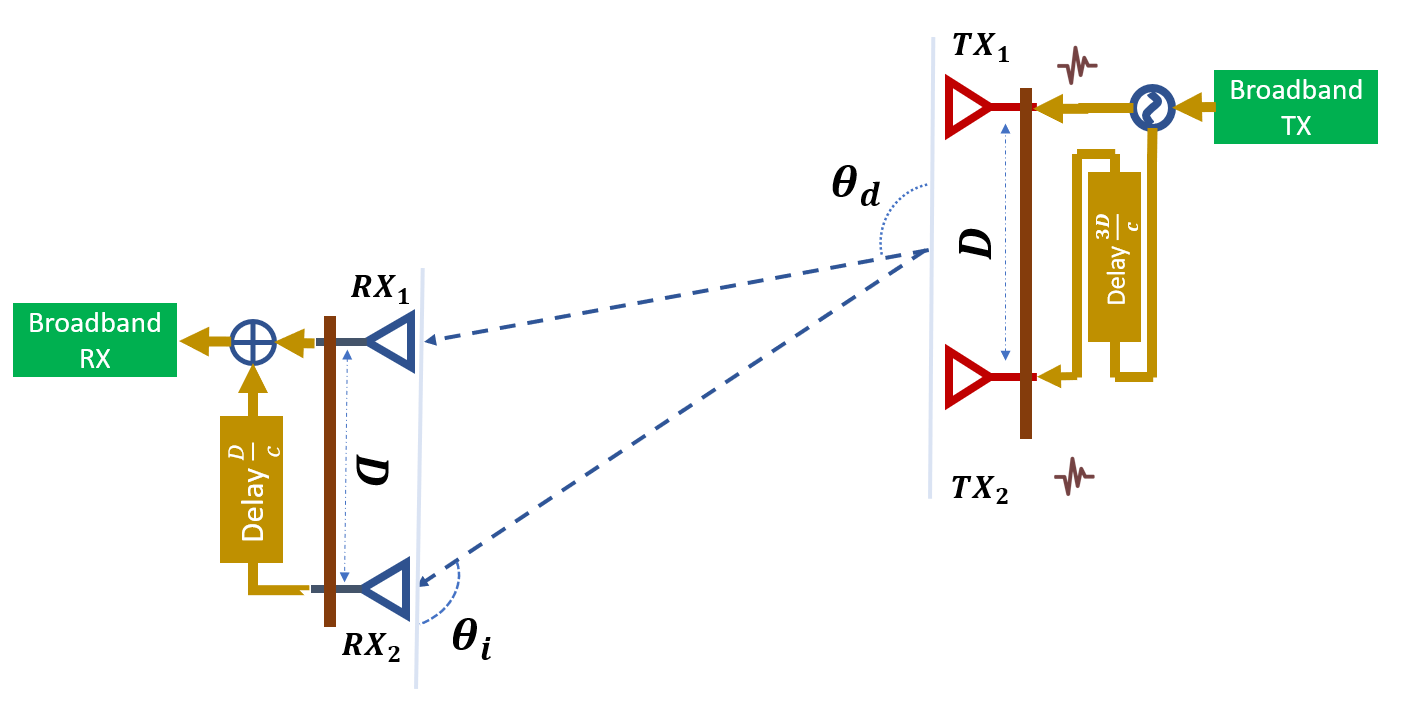}
    \caption{Block diagram of spectrum shaping for estimating AoD and DoA together. The TX antenna configuration is almost identical to the RX unless the delay line before $TX_2$ is 3 times longer than the delay line after $RX_2$.} 
    \label{BD}
\end{figure}

THz rainbow \cite{ghasempour2020single} introduces a technique to measure AoD of signal from the TX. THz rainbow takes advantage of the derivative of the maximum and the minimum received frequency with respect to the RX rotation to estimate AoD. Here, we introduce a scheme that can be used for AoD estimation in a single shot without any reliance of RX rotation leveraging a TX antenna configuration similar to the RX as illustrated in Fig. \ref{BD}. 

Assuming there is only LoS path between the TX and the RX, we propose using the same antenna placement of the RX at the TX. There is a delay line between the source and $TX_2$ with length $3D$. Thus, $TX_2$ propagates the pulse generated by the source with a delay equal to $\frac{3D}{c}$ with respect to the $TX_1$ emission. The two pulses will be received at the $RX_1$ and $RX_2$ with a delay due to the AoD and the delay between the transmission of the two pulses equals to 
\begin{equation}
 \delta_d =- \frac{D}{c}\cos(\theta_d) + \frac{3D}{c} \,, 
 \end{equation}
where $\delta_d$ denotes the difference time of arrival of the two transmitted pulses at $RX_1$, and $\theta_d$ denotes AoD. Denoting the emitted pulse from $TX_1$ by $s(t)$, the received signal at $RX_1$ turns out to be 

\begin{align}
\label{rsig}
    &r(t) = a(t) *\left(s(t) + s(t-\frac{D}{c}(3-\cos(\theta_d))\right)\\ \nonumber
    &\Rightarrow R(f) = \mathcal{F}\{r(t)\} =  a(f)S(f)(1+e^{-j2\pi f \frac{D}{c}(3-\cos(\theta_d)})=a(f)S(f)e^{-j\pi f \frac{D}{c}(3-\cos(\theta_d))}\cos(\pi f \frac{D}{c}(3-\cos(\theta_d)))\,.
\end{align}

Using \eqref{spes}, $\boldsymbol{E}_r(f,\theta_i,\theta_d)$ can be expressed as

\begin{align}
    \label{specsp1}
    \boldsymbol{E}_r(f,\theta_i,\theta_d) =  |4a(f)S(f)|^2\left|\cos(\pi f \frac{D}{c}(1-\cos(\theta_i)))\cos(\pi f \frac{D}{c}(3-\cos(\theta_d)))\right|^2 \\ \nonumber
    = |2a(f)S(f)|^2\left(\left(1+\cos(2\pi f \frac{D}{c}(1-\cos(\theta_i)))\right)\left(1+\cos(2\pi f \frac{D}{c}(3-\cos(\theta_d)))\right) \right)
    \\ \nonumber
    = |2a(f)S(f)|^2 ( 1 + \cos(2\pi f \frac{D}{c}(1-\cos(\theta_i))) + \cos(2\pi f \frac{D}{c}(3-\cos(\theta_d))) \\ \nonumber
    + \frac{1}{2} \cos(2\pi f \frac{D}{c}(4-\cos(\theta_i)-\cos(\theta_d))) +\frac{1}{2} \cos(2\pi f \frac{D}{c}(2-\cos(\theta_d)+\cos(\theta_i)))) 
    \,.
\end{align}

Based on \eqref{specsp1}, $\boldsymbol{E}_r(f,\theta_i,\theta_d)$ is a summation of four harmonics with frequencies equal to $\frac{D}{c}(1-\cos(\theta_i))$, $\frac{D}{c}(3-\cos(\theta_d))$, $\frac{D}{c}(4-\cos(\theta_i)-\cos(\theta_d))$, and $\frac{D}{c}(2-\cos(\theta_d)+\cos(\theta_i))$.

$\boldsymbol{Lemma\: 2.}$ For any AOA $\theta_i$ and AOD $\theta_d$ we have 
\begin{align}
   \frac{D}{c}(4-\cos(\theta_i)-\cos(\theta_d)) \ge  \frac{D}{c}(3-\cos(\theta_d)) \ge \frac{D}{c}(1-\cos(\theta_i)), \frac{D}{c}(2-\cos(\theta_d)+\cos(\theta_i)) \ge 0 \,.
\end{align}

\textbf{Proof.}  Assume $\alpha = \frac{D}{c}(3-\cos(\theta_d))$ and $\beta = \frac{D}{c}(1-\cos(\theta_i))$, $\forall{\theta_i},\forall{\theta_j}$, $\frac{4D}{c}\ge\alpha\ge \frac{2D}{c}$ and $\frac{2D}{c}\ge\beta\ge 0$. Thus we have 
$$\alpha \ge \beta \ge 0.$$Therefore $$\alpha + \beta \ge \alpha \ge \beta, \alpha- \beta \ge 0.$$ 
Hence, \textbf{Lemma 2} holds.$\square$ 

{ Applying Fourier transform on $\boldsymbol{E}_r(f,\theta_i,\theta_d)$ similar to \eqref{arres2}, the four frequencies of the four harmonics of the $\boldsymbol{E}_r(f,\theta_i,\theta_d)$ are measurable. Referring to \textbf{Lemma 2}, the two largest frequencies of $\boldsymbol{E}_r(f,\theta_i,\theta_d)$ are  $\frac{D}{c}(4-\cos(\theta_i)-\cos(\theta_d))$ and $\frac{D}{c}(3-\cos(\theta_d))$, respectively. Using $\frac{D}{c}(3-\cos(\theta_d))$, $\theta_d$ can be uniquely estimated if $\theta_d \in [0,\pi]$. Next, using $\frac{D}{c}(4-\cos(\theta_i)-\cos(\theta_d))$, $\theta_i$ can be uniquely estimated if $\theta_i \in [0,\pi]$. Referring to \cite{ghasempour2020single}, using THz rainbow technique $\theta_d$ is measurable in a very limited range ($[\theta_i-40^o,\theta_i+20^o]$). However, leveraging the SSH introduced in this section $\theta_d$ is measurable in the whole range of $0^o$ to $180^o$. Referring to \eqref{specsp1}, here we applied SSH both at the RX (the $\cos(\pi f \frac{D}{c}(1-\cos(\theta_i)))$ term) and the TX (the $\cos(\pi f \frac{D}{c}(3-\cos(\theta_d)))$ term) to have enough information to estimate both AoD and DoA. In this work, we have mainly made use of two antennas and one delay line to from spectrum shapers. Nonetheless, future work can go further by introducing more complex antenna and delay line schemes for more advanced applications and superior performances.      
\section{Cramer Rao Lower Bound of Error}
\label{CRBSec}
In this section, we analyze the effect of noise on the performance of the system. Specifically, we derive Cramer Rao lower bound (CRB) of error for DoA estimation for LoS scenario (Section \ref{secA}) and CRB for AoD and DoA estimation (Section \ref{DoAAoD}). 
\subsection{CRB of DoA Estimation}
Referring to Section \ref{secA}, we assume received signals at $RX_1$ and $RX_2$ are added by a white Gaussian noise with noise power $\frac{N_0}{2}$. We have 
\begin{align}
   &r_1(t) = a(t)*s(t) + n_1(t) \nonumber \\
    &r_2(t) = a(t)*s(t-\delta t_i) +n_2(t)
\end{align}
where $n_1(t)$ and $n_2(t)$ are independent white Gaussian noises with power $\frac{N_0}{2}$. 
The received signals at the THz-TDS receiver equals to
\begin{align}
 &r_1(t)+r_2(t-\frac{D}{c})=a(t)*(s(t)+s(t-\frac{D}{c}-\delta t_i))+n_1(t)+n_2(t) \,.
 \label{obs}
\end{align}
 We define $n(t)=n_1(t)+n_2(t)$ which is a white Gaussian noise with power $N_0$. The observation (denoted by $z(f)$) is the spectrum of the received signal equal to
\begin{align}
 &z(f) = |a(f)s(f)\left(1+e^{j2\pi f \frac{D}{c}(1-\cos(\theta_i))}\right)+n(f)|\,.
 \label{obs1}
\end{align}

$n(f)$ is a white Gaussian noise process with power $N_0$ and $\boldsymbol{x}=(\theta_i,a(f))$ denotes the vector of unknowns. Referring to \cite{papoulis1989probability}, the likelihood function of the observation is a Rice distribution with the following parameters
\begin{align}
p(z(f);\boldsymbol{x})= \frac{z(f)}{\frac{N_0}{2}}e^{-\left(\frac{z(f)^2+4a(f)^2s(f)^2\cos^2(\pi f \frac{2D}{c}\sin^2(\frac{\theta_i}{2}))}{N_0}\right)}I_0(\frac{2z(f)|a(f)s(f)\cos(\pi f \frac{2D}{c}\sin^2(\frac{\theta_i}{2}))|}{\frac{N_0}{2}})
\label{likilihood}
\end{align}
where $I_0()$ is the modified Bessel function of the first kind with order zero, and $p(z(f);\boldsymbol{x})$ is the probability distribution function of $z(f)$ 
parametrized by the $\boldsymbol{x}$. The covariance matrix of any unbiased estimator $\hat{\boldsymbol{x}}$ is bounded by \cite{chepuri2014sparsity}
\begin{align}
    E\{(\hat{\boldsymbol{x}}-{\boldsymbol{x}})(\hat{\boldsymbol{x}}-{\boldsymbol{x}})^H\} \ge \boldsymbol{J}^{-1}\,,
\end{align}
where 
\begin{align}
    \boldsymbol{J}({\boldsymbol{x}}) = \sum_{k = 1}^{M_f} E \left\{ \left(\frac{\partial \ln{p(z(f_k);\boldsymbol{x})}}{\partial\boldsymbol{x}}\right)\left(\frac{\partial \ln{p(z(f_k);\boldsymbol{x})}}{\partial\boldsymbol{x}}\right)^T\right\} \,.
    \label{Jrev}
\end{align}
Here $M_f$ denotes the number of frequency samples of $z(f)$ in the relevant band, and $f_1,\dots,f_{M_f}$ are the frequencies at which $z(f)$ is sampled by the THz-TDS technique. The analytical closed-form derivation of $\boldsymbol{J}$ and consequently CRB for $\theta_i$ estimation is not feasible because of the Bessel function. In Section \ref{sim} we will resort to a numerical approximation of CRB for the sake of noise analysis of SSH  and comparison with the state-of-the-art.

\subsection{CRB of AoD Estimation}
Based on the discussion in Section \ref{DoAAoD}, referring to \eqref{rsig}, the observation at the RX would be in the form of 
\begin{align}
z(f)=|a(f)s(f)\left((1+e^{j2\pi f \frac{D}{c}(1-\cos(\theta_i))})(1+e^{j2\pi f\frac{D}{c}(3-\cos(\theta_d)})\right)+n(f)|\,.
\end{align}
Defining $\boldsymbol{x}=(\theta_i,\theta_d,a(f))$ as the vector of unknown parameters, and given $n(f)$ is white Gaussian noise, the likelihood function turns out to be
\begin{align}
p(z(f);\boldsymbol{x})= \frac{z(f)}{\frac{N_0}{2}}e^{-\left(\frac{z(f)^2+16a(f)^2s(f)^2\cos^2(\pi f \frac{2D}{c}\sin^2(\frac{\theta_i}{2}))\cos^2(\pi f \frac{2D}{c}(2+\sin^2(\frac{\theta_d}{2})))}{N_0}\right)} \times \nonumber \\
I_0\left(\frac{4z(f)|a(f)s(f)\cos(\pi f \frac{2D}{c}\sin^2(\frac{\theta_i}{2}))\cos(\pi f \frac{2D}{c}(2+\sin^2(\frac{\theta_d}{2})))|}{\frac{N_0}{2}}\right) \,,
\label{likilihood1}
\end{align}

where $\times$ is multiplication.
Then using \eqref{Jrev}, $\boldsymbol{J}$ can be calculated and CRB for $\theta_i$ and $\theta_d$ can be derived using numerical techniques.}
\section{Simulation Results}
\label{sim}
{
\subsection{Noise Analysis}
In this section, the performance of our proposed link discovery technique is studied in the presence of noise. Further, the performance of the proposed technique is compared with the performance of the state-of-the-art lens array and ULA using digital beamforming. The performance of ULA using digital beamforming has been extensively studied in the literature \cite{ntouni2020real,busari2019generalized}. Implementing digital beamforming for ULA is exorbitant and massively complex especially in mmwave bands \cite{kutty2015beamforming}. On the other hand, Lens array (LA) is a more recent technique utilizes a lens as a passive phase shifter to pursue beamforming without the heavy network of phase shifters required in ULA arrays \cite{cho2018rf}. Throughout this section, signal to noise ratio (SNR) is referred to the average power of signal to noise variance ($\frac{N_0}{2}$) at each antenna of the array (whether ULA, LA, or SSH).  

In the first simulation, we compare CRB of DoA estimation for a symmetric ULA with half wavelength antenna spacing equipped with $N$ elements \cite{penna2011bounds}, an arc LA with aperture and focal length equals to $L$ equipped with $M$ elements \cite{shim2018cramer}. Further we set $L = M\frac{\lambda}{2}$, where $\lambda$ is the wavelength of the operating frequency of the LA. To emulate SSH, we assume the spectral resolution of the THz-TDS system is 1.5 GHz,  and the pulse relevant band is [100GHz,1THz]. Thus the number of spectrum samples are 600. We set $D = 5$ mm. For the channel frequency response $a(f)$ we assume the range is 10 0m and the air is dry\footnote{We simulate signal attenuation due to atmospheric gases using MATLAB$^\textsuperscript{\textregistered}$ \texttt{gaspl} function.} (unless otherwise mentioned). As Fig. \ref{CRBmain} illustrates CRB of SSH is close to a ULA with 111 elements and LA with 201 elements when SNR is 20 dB, which is a spectacular performance considering much simpler hardware architecture of SSH compared to the two other counterparts. Regarding Fig. \ref{CRBmain} (a) to (d), when SNR decreases the performance of SSH deteriorates sharper than ULA and LA in the CRB sense. Specifically, looking at Fig. \ref{CRBmain} (d), when SNR = -10 dB, the CRB of SSH is in proximity of a ULA with 7 elements and a LA with 15 elements. 
This performance decline is due to the fact that large arrays can integrate the received signal energy over the whole antenna elements, while SSH only utilizes 2 antennas, thus does not benefit from the same advantage. Speaking of hardware complexity, even in low SNR levels, SSH still has superior performance compared to ULA and LA. As shown, SSH shows identical performance to a far larger ULA and LA.

\begin{figure}
\begin{subfigure}{.5\textwidth}
  \centering
  \includegraphics[width=3.2in,height=2.2in]{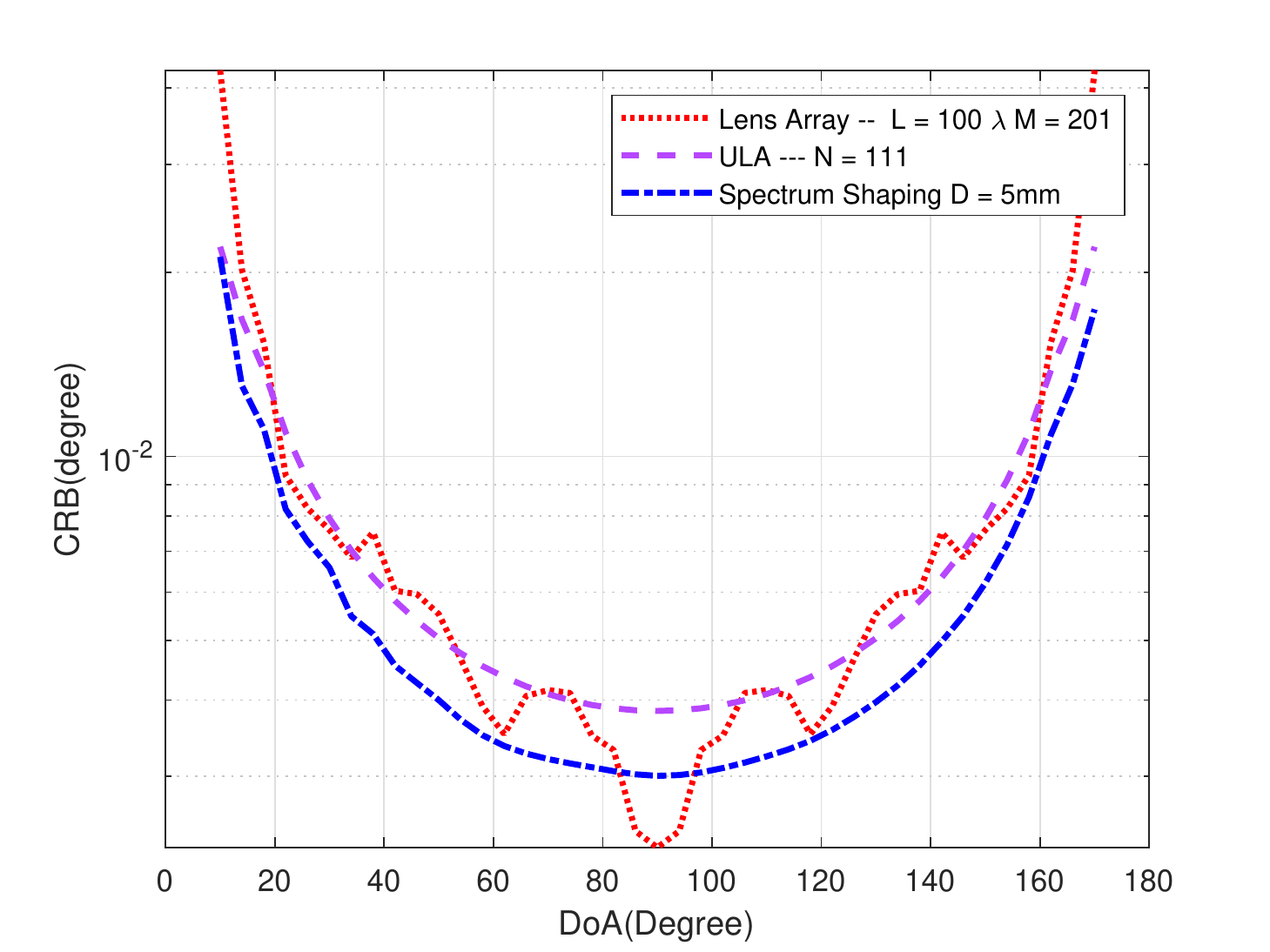}
  \caption{SNR = 20dB}
  \label{fig:sfig1}
\end{subfigure}%
\begin{subfigure}{.5\textwidth}
  \centering
  \includegraphics[width=3.2in,height=2.2in]{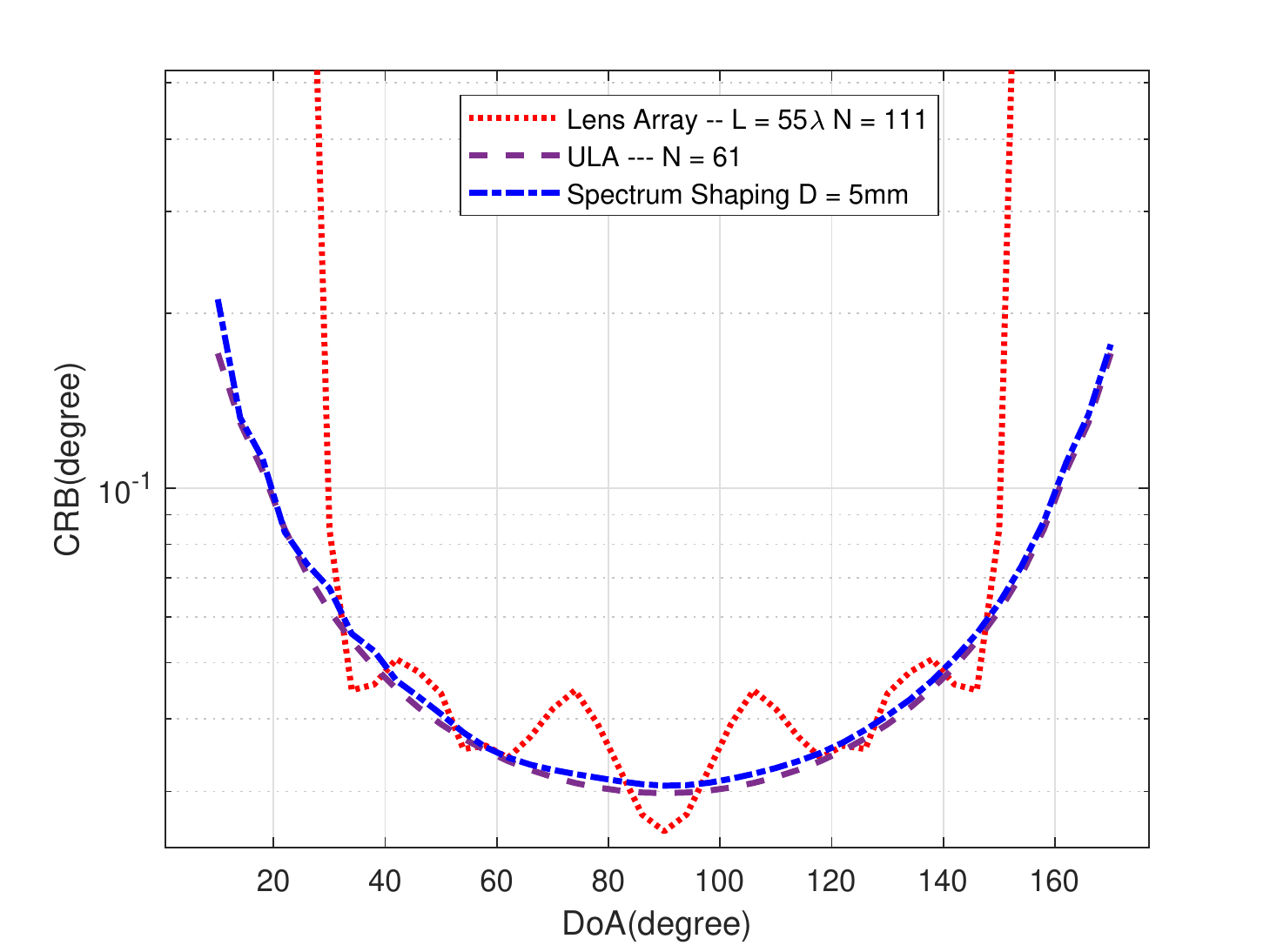}
  \caption{SNR = 10dB}
  \label{fig:sfig2}
\end{subfigure}
\begin{subfigure}{.5\textwidth}
  \centering
  \includegraphics[width=3.2in,height=2.2in]{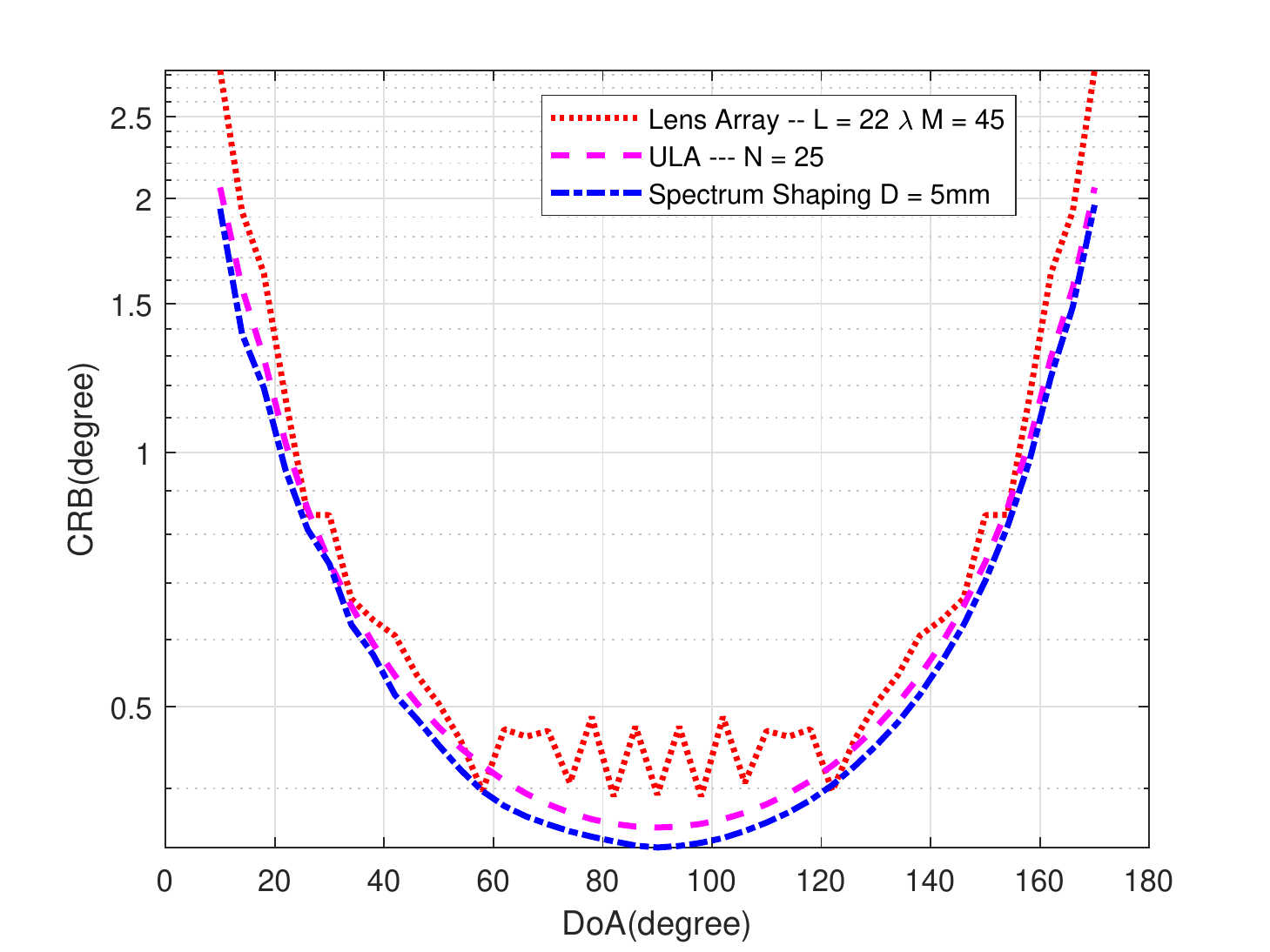}
  \caption{SNR = 0dB}
  \label{fig:sfig3}
\end{subfigure}
\begin{subfigure}{.5\textwidth}
  \centering
  \includegraphics[width=3.2in,height=2.2in]{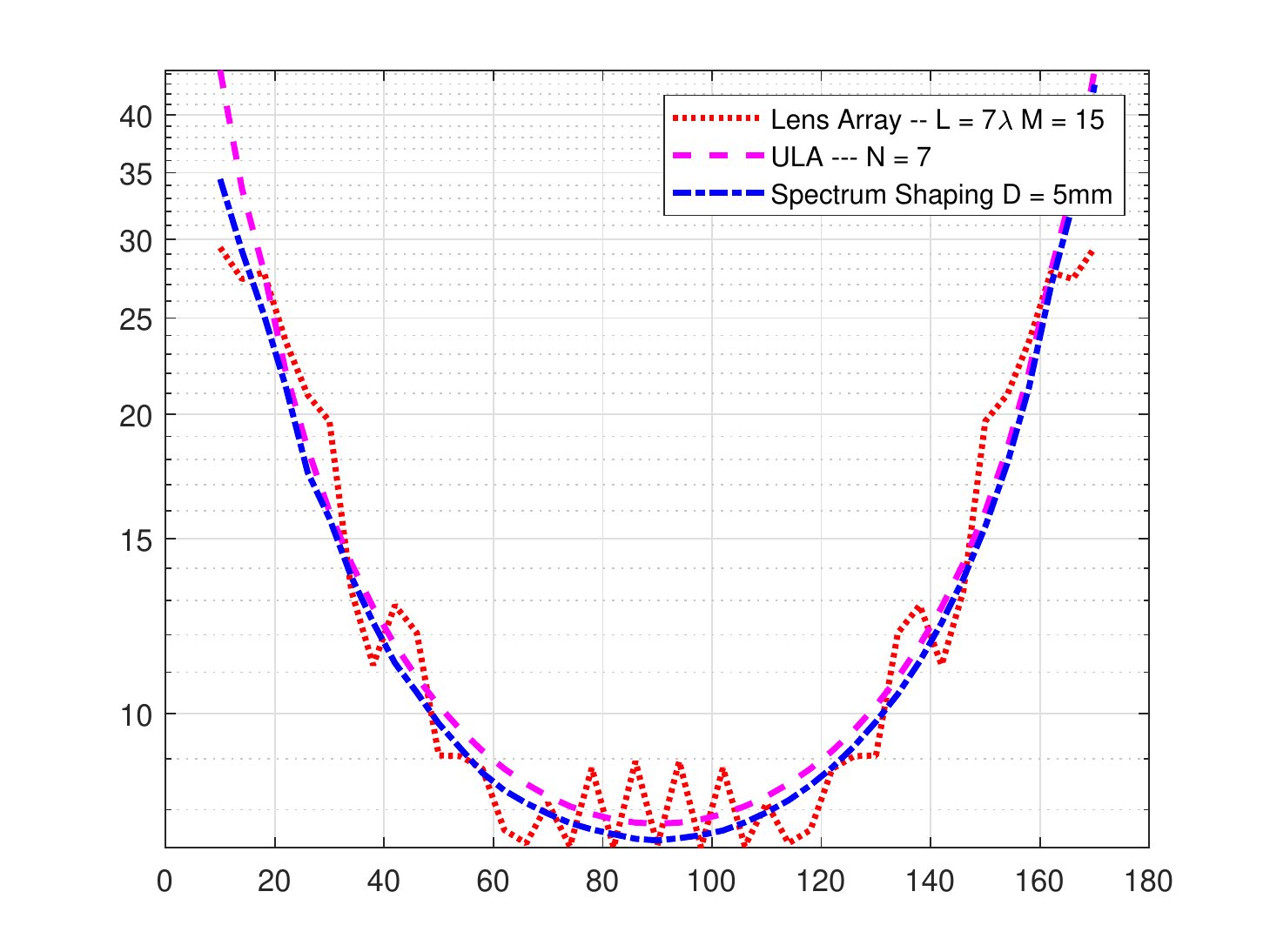}
  \caption{SNR = -10dB}
  \label{fig:sfig3}
\end{subfigure}
\caption{CRB of DoA estimation is compared to  ULA, Lens array and spectrum shaping techniques. Here $\lambda$ denotes wavelength.}
\label{CRBmain}
\end{figure}

In the next simulation we investigate the effect of harsh frequency selective channel on the performance of the CRB of the SSH technique. The THz channel can be extremely frequency selective because of attenuation of atmospheric gases specially water vapor and this effect exacerbates as range increases \cite{gaspl}. Here we assume that the weather is exceedingly humid and water vapor density equals $10 \frac{g}{m^3}$. We set SNR = 5 dB and D = 5 mm. Fig. \ref{CRBWV} shows CRB of SSH for ranges equal to 10 m, 100 m, and 1000 m. As expected, as range increases the SSH performance deteriorates, since a large range of frequencies in the relevant bandwidth experiences severe attenuation. Nevertheless, SSH still can achieve DoA estimation accuracy of better than $2^o$ for DoA $\in [30^o,145^o]$ for range equal to 1 km and in the presence of intense water vapor, which quite acceptable for majority of directional communication applications.

\begin{figure}
  \centering
  \includegraphics[width=4.2in,height=3.2in]{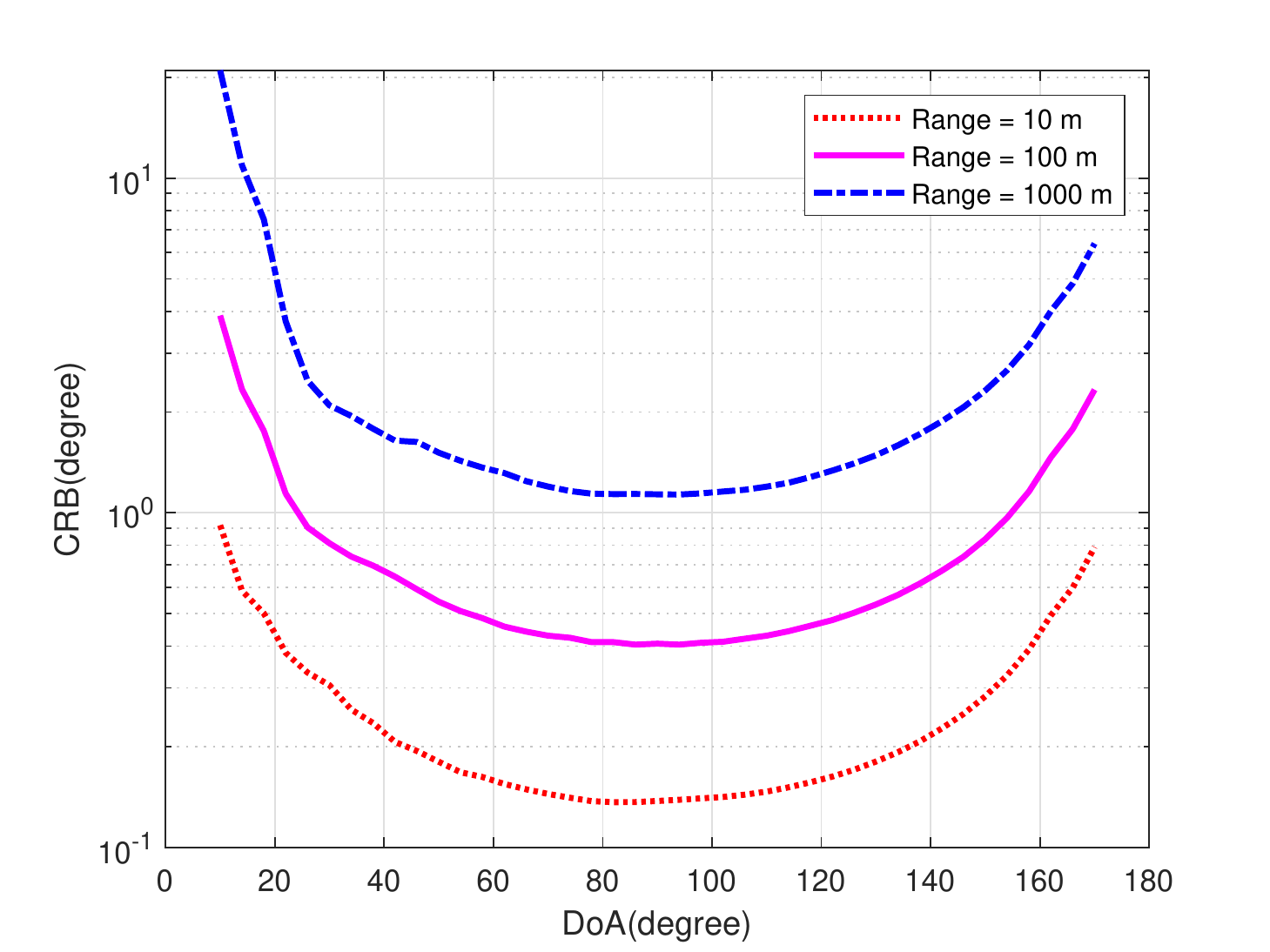}
  \caption{CRB of spectrum shaping in the presence of intensive water vapor in the air for different ranges. Here $D =5 mm$, SNR = 5 dB. }
  \label{CRBWV}
\end{figure}

In the next simulation we study the effect of $D$, the antenna gap, on CRB of the SSH technique. In this simulation we assume SNR = 0 dB. We simulate CRB for D equals to 1 mm, 5 mm, and 10 mm. As Fig. \ref{CRBD} shows, SSH performs more accurately as $D$ increases. Referring to \eqref{arres2}, $D$ directly multiplies to $f\cos(\theta_i)$, thus increasing $D$ is equivalent of increasing the relevant band. Therefore $D$ is one the major parameter can improve SSH performance. 

\begin{figure}
  \centering
  \includegraphics[width=4.2in,height=3.2in]{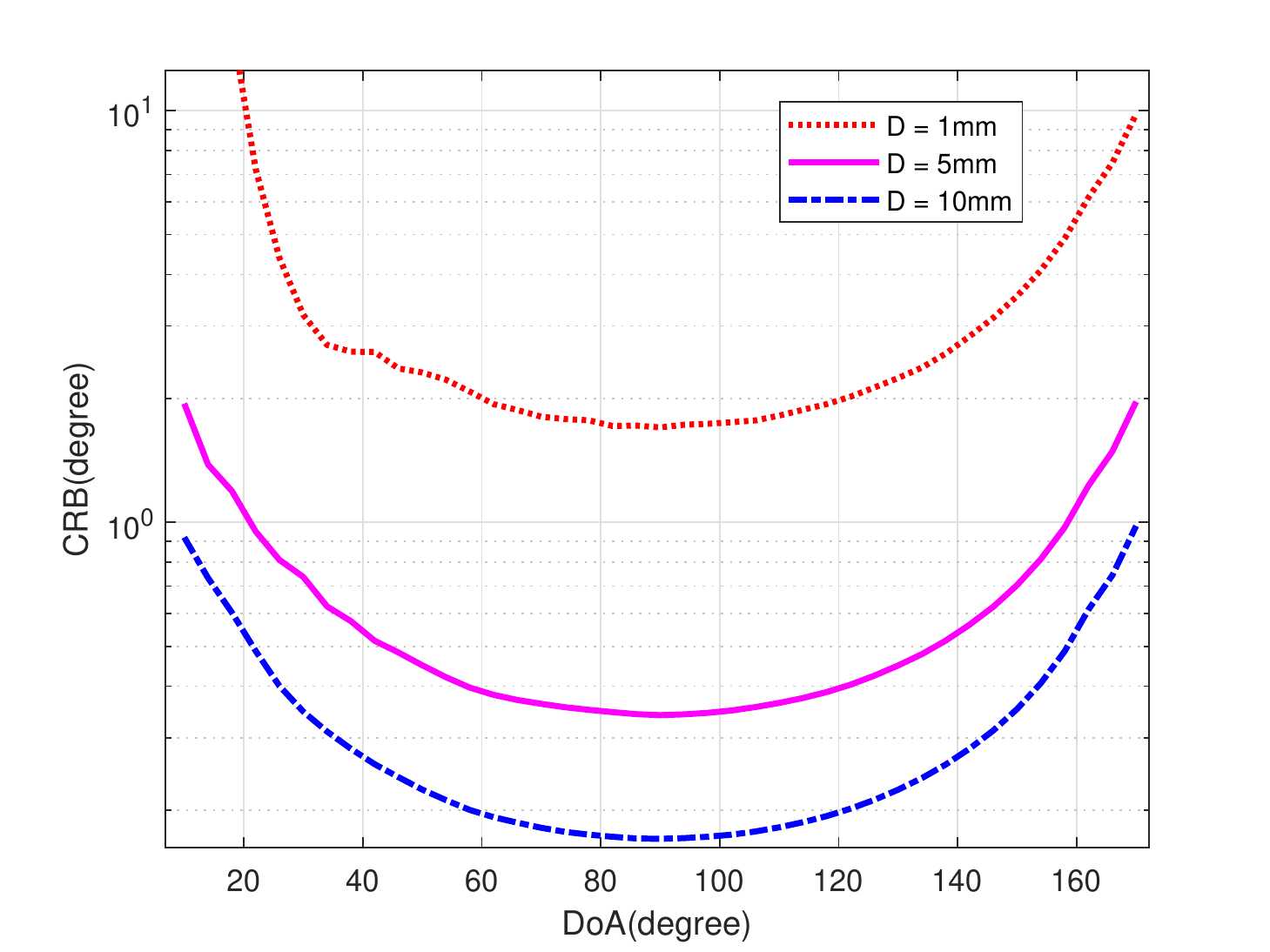}
  \caption{CRB of spectrum shaping for different values of antenna spacing $D$. Here SNR = 0dB. }
  \label{CRBD}
\end{figure}

The next simulation studies the CRB of AoD and DoA estimation of Section \ref{DoAAoD}. Here we assume SNR = 5 dB, and $D = 5$ mm. Fig. \ref{CRBAODDOA} demonstrates that AoD and DoA have almost identical CRBs. Further, AoD has negligible effect on CRB of DoA (which means different values of AoD results in the same DoA CRB) and vise versa.  

\begin{figure}
\begin{subfigure}{.5\textwidth}
  \centering
  \includegraphics[width=3.2in,height=2.2in]{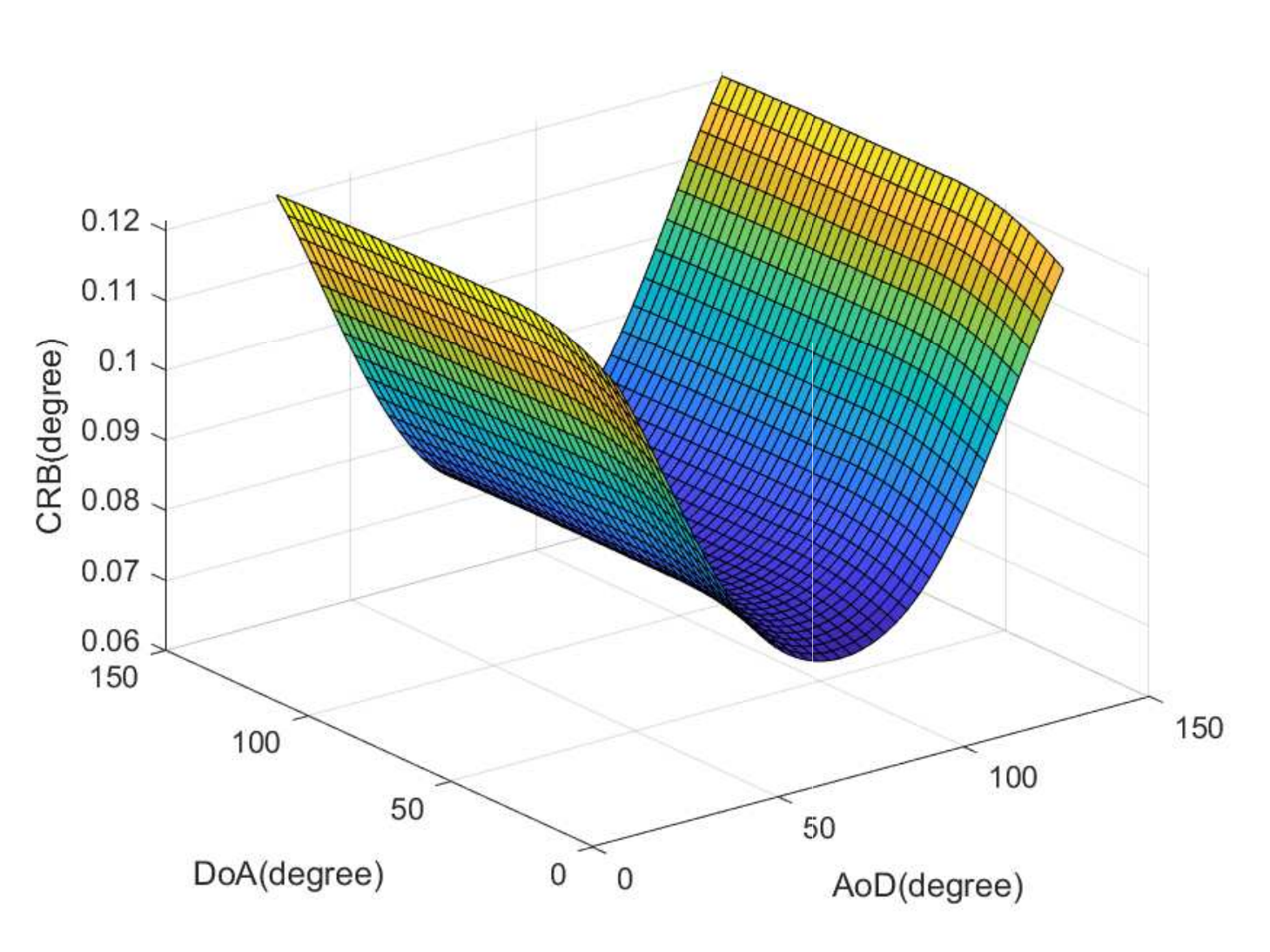}
  \caption{AoD}
  \label{figAoD}
\end{subfigure}%
\begin{subfigure}{.5\textwidth}
  \centering
  \includegraphics[width=3.2in,height=2.2in]{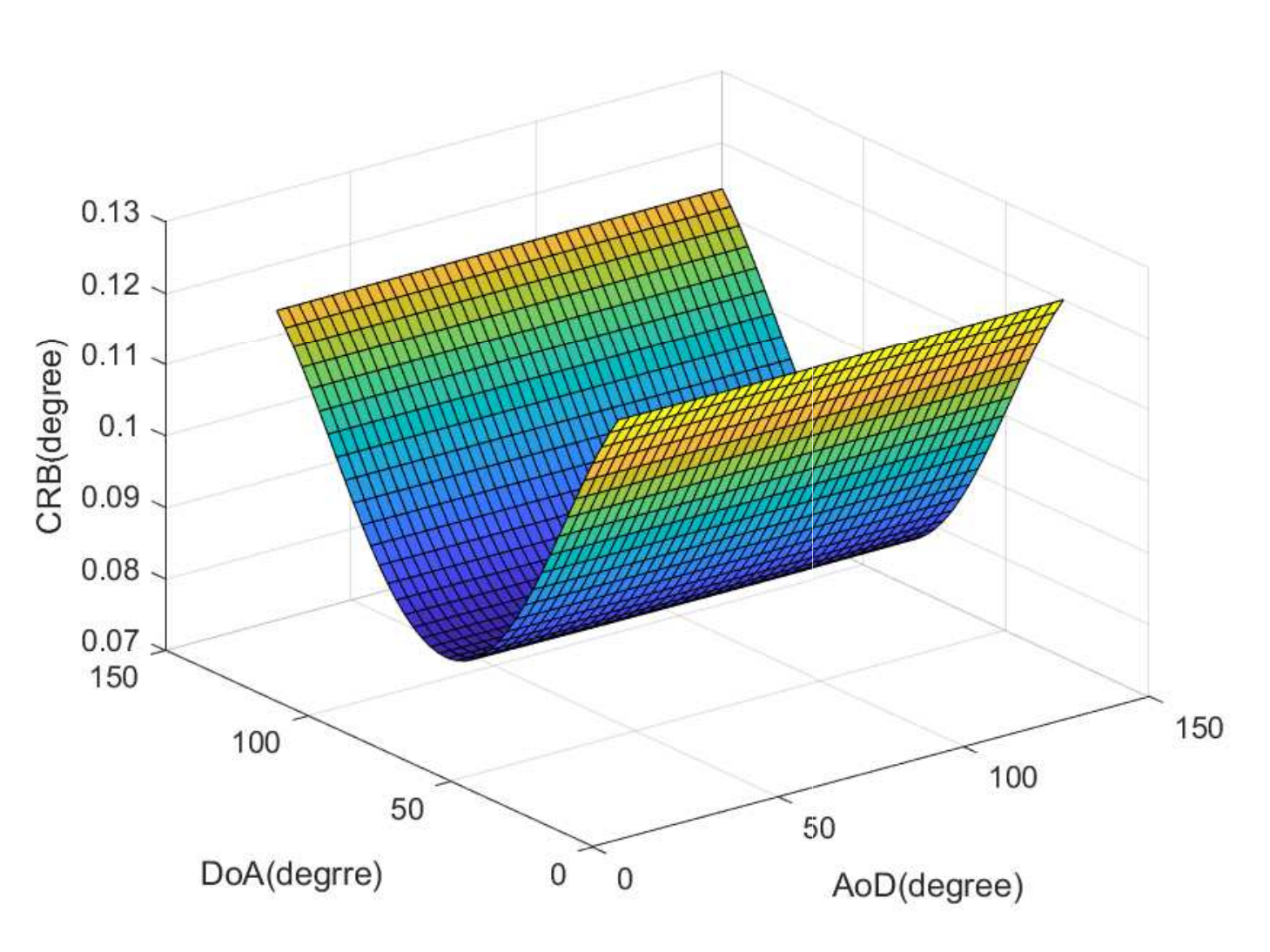}
  \caption{DoA}
  \label{figDoA}
\end{subfigure}
\caption{CRB of AoD and DoA estimation considering antenna configuration of Fig. \ref{BD}. Here $D = 5 mm$ and SNR = 5 dB.}
\label{CRBAODDOA}
\end{figure}

In the next simulation the mean squared error (MSE) of minimum mean square estimator (MMSE) for ULA, LA and SSH is simulated and compared. Assuming array response in a noisy environment is denoted by $A(\theta)$ and the expected response in an ideal noise-free environment is denoted by $A_E(\theta)$, \NR{rewrite this sentence}for SSH, $A_E(\theta)$ equals $\boldsymbol{E}_r(f,\theta_i)$, the MMSE estimator can be defined as 
$$ \hat{\theta} = \min_{0^o\le\theta \le 180^o} |A(\theta)-A_E(\theta)|^2 \,,$$
where $\theta$ is DoA and $\hat{\theta}$ is the estimated DoA. Thus the MSE can be defined as the average of squared error  ($|\hat{\theta}-\theta_0|^2$
, where $\theta_0$ is the actual DoA). In this simulation we compare the performance of a ULA with N = 60, an arc LA with M = 80 and SSH with $D = 5 mm$. To calculate MSE we repeat the experiment for 1000 times and average over all experiments' results. As Fig. \ref{ErrorComp} illustrates ULA shows lower root MSE (RMSE) for negative SNRs in comparison with SSH and LA. SSH performs slightly better than ULA and LA for SNR larger than 3 dB. These results is in harmony with the outcome of CRB analysis that SSH performance improves significantly as SNR increases.

\begin{figure}
  \centering
  \includegraphics[width=4.2in,height=3.2in]{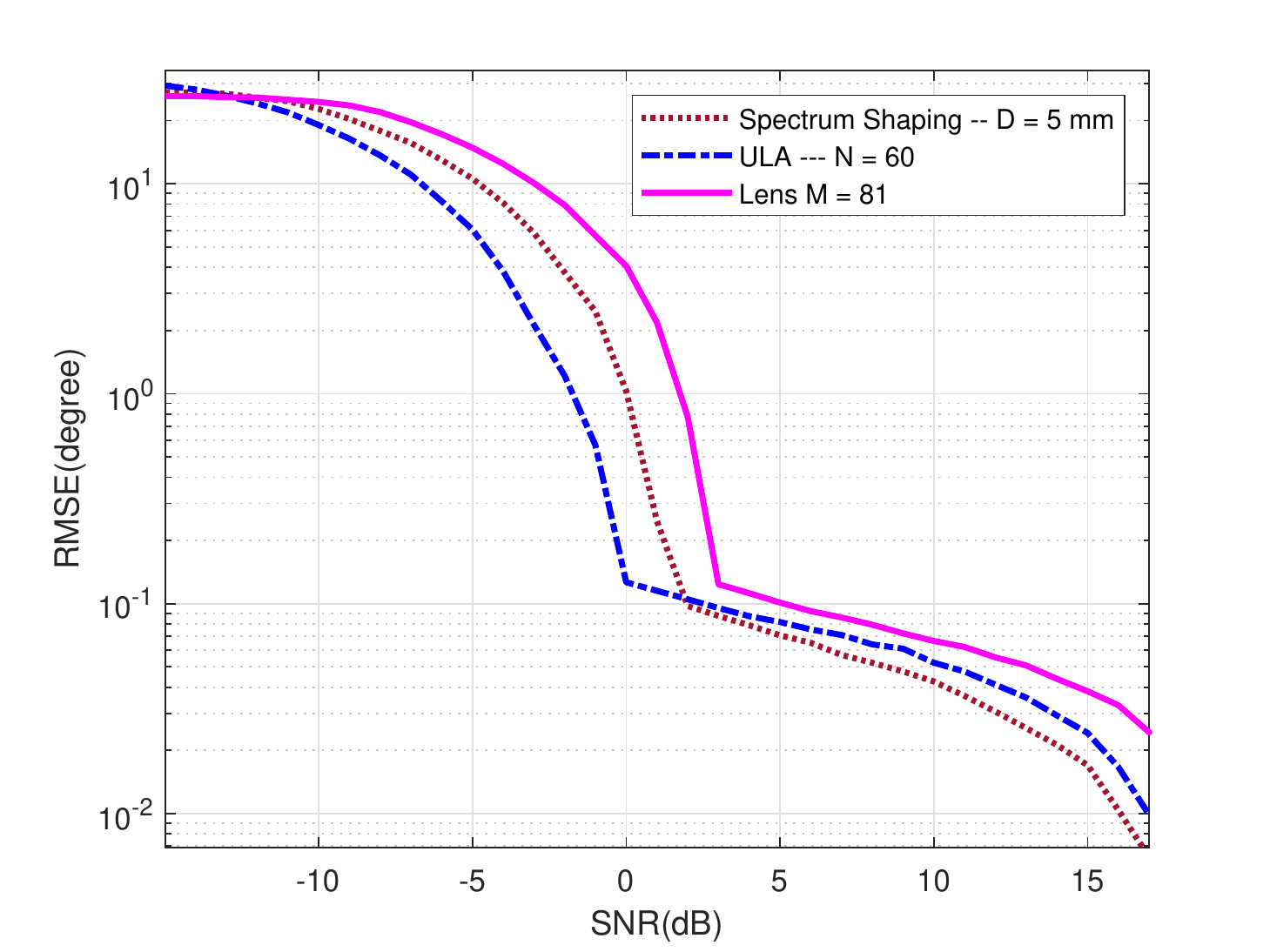}
  \caption{RMSE of ULA, LA and spectrum shaping versus SNR. Spectrum shaping performance surpasses large arrays in positive SNR, while spectrum shaping attains a much simpler hardware structure in comparison to arrays.  }
  \label{ErrorComp}
\end{figure}

Next, we study RMSE of SSH for different antenna spacing in the presence of intense water vapor. We set SNR = 5 dB, DoA = $60^o$, and water vapor density to $10 \frac{g}{m^3}$. As Fig. \ref{EWV} demonstrates SSH with wider antenna spacing performs significantly better in the presence of intense water vapor. Nonetheless, as range increases the performance of SSH regardless of antenna spacing deterioration due to  the harsh frequency selective attenuation of the channel. Additionally, even in the presence of expensively harsh frequency selective channel, SSH with 5 mm antenna spacing can achieve better than $3^o$ accuracy at the range of 1800 m.        

\begin{figure}
  \centering
  \includegraphics[width=4.2in,height=3.2in]{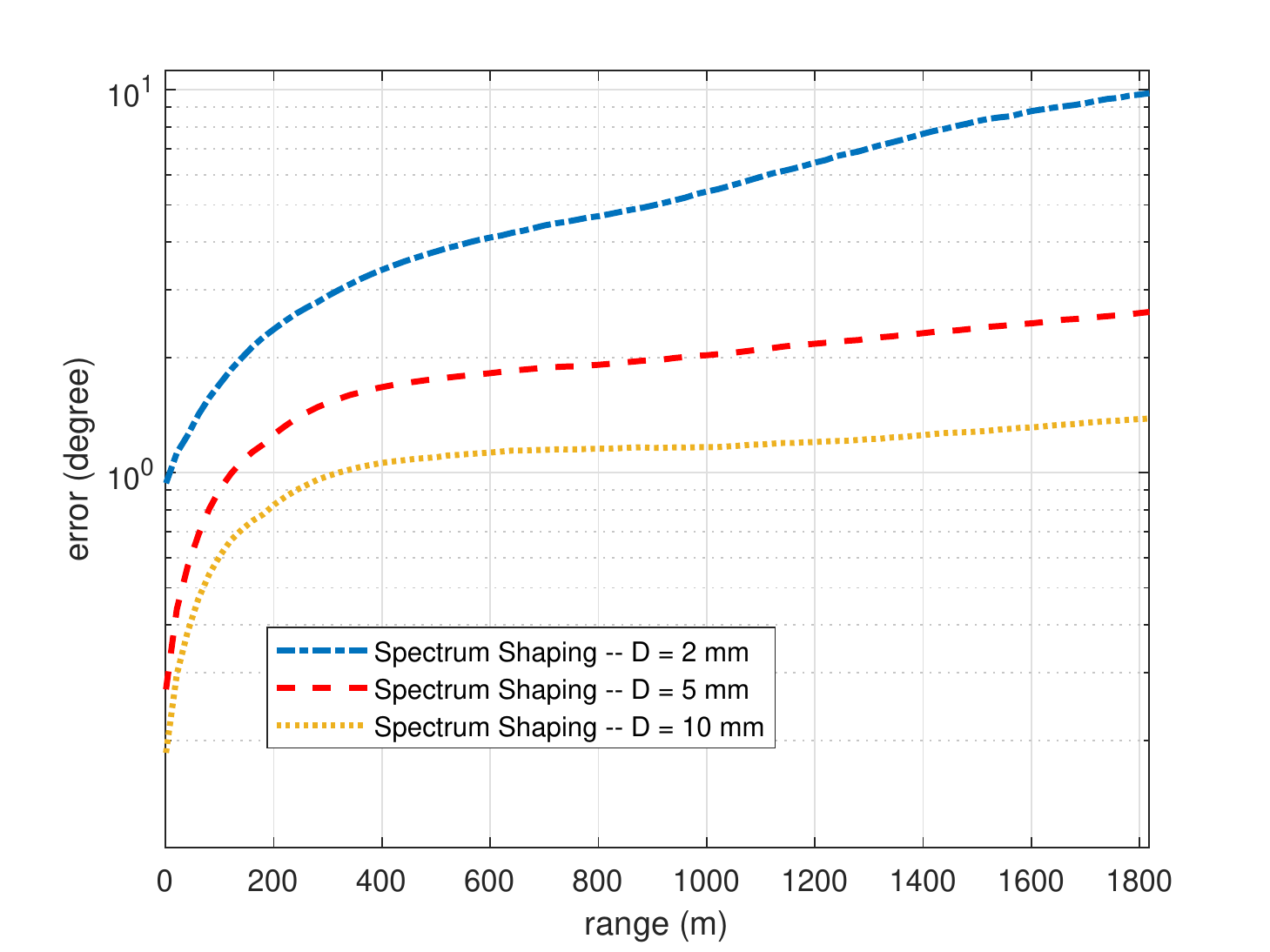}
  \caption{RMSE of DoA estimation in the presence of intense water vapor ($10    \frac{g}{m^3}$) } versus range assuming SNR stays intact (5 dB). DoA equals $60^o$. 
  \label{EWV}
\end{figure}

In the next experiment, we examine the performance of SSH in simultaneous AoD and DoA estimation. We set D = 5 mm, and SNR = 5 dB. As Fig. \ref{AD} illustrates and as expected, the RMSE of both AoD and DoA are mostly identical and improve remarkably as SNR increases. This identical response is due to the almost identical configuration at the TX and the RX and is in agreement with the result of CRB analysis.  

\begin{figure}
  \centering
  \includegraphics[width=4.2in,height=3.2in]{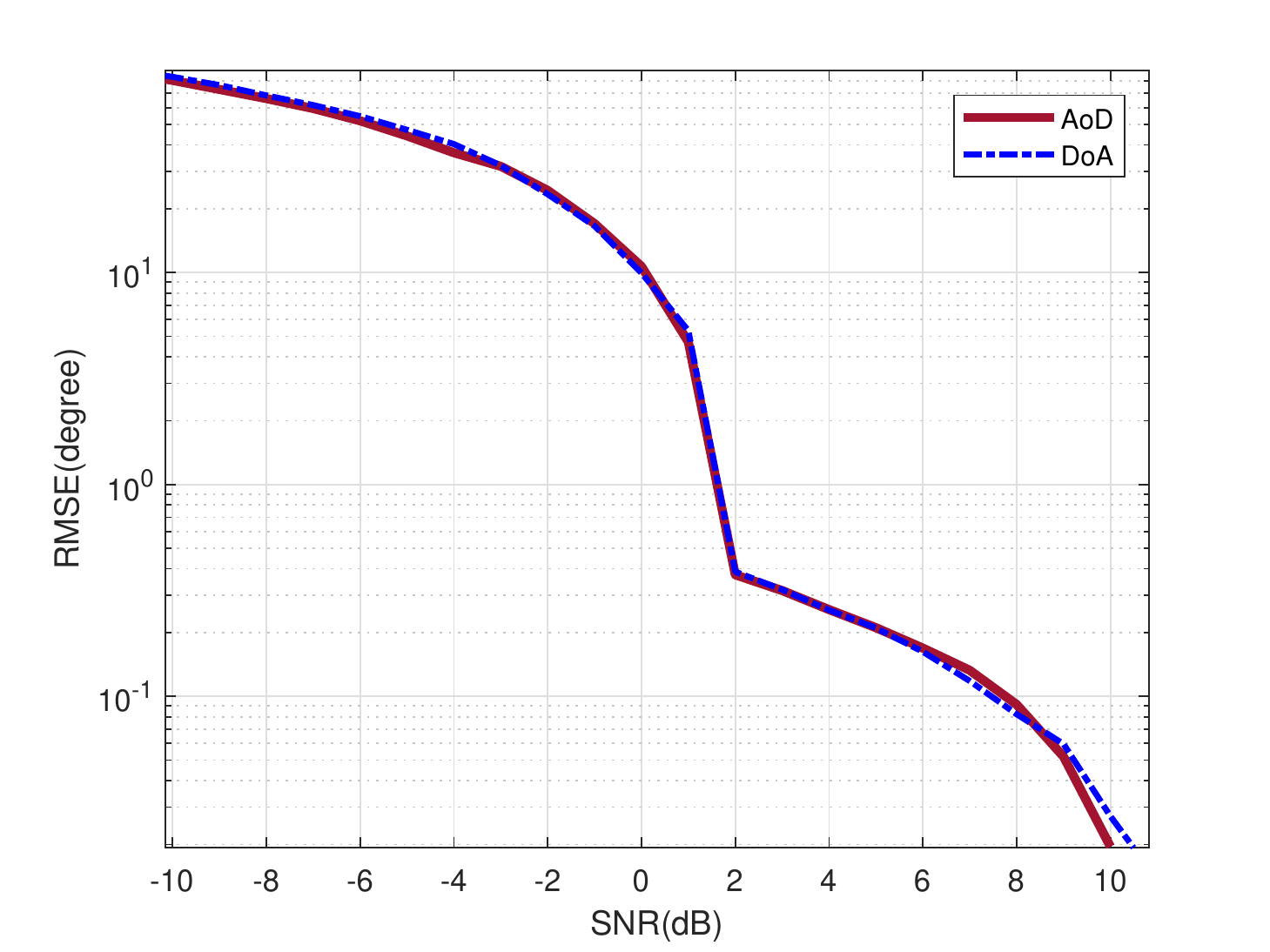}
  \caption{RMSE of DoA and AoD estimation using MMSE estimator. DoA equals $60^o$ and SNR = 5 dB.} 
  \label{AD}
\end{figure}

\subsection{Channel State Information Estimation Using Spectrum Shaping}
Here we consider an indoor scenario where there are two paths between the TX and the RX, one LoS and one NLoS. We set the DoA of the LoS path $60^o$ and the DoA of NLoS path $100^o$. We assume the NLoS path is $0.5$ m longer than the LoS path and the power of NLoS path is 6 dB less than that of the LoS path. We suppose the TX is transmitting a quasi sinc pulse with flat bandwidth between [0.1THz,1THz]. We set $D = 5mm$. We suppose  To find the DoAs of incoming signals we simply apply the matched filter of the expected harmonic (derived in \eqref{arres1}) on  the output of THz-TDS ($\boldsymbol{E}_r(f,\theta_1,\theta_2)$) to form $\boldsymbol{E}(\theta)$ given by
\begin{equation}
    \boldsymbol{E}(\theta) = \int_{0.1THz}^{1THz} \cos(2\pi f \frac{2D}{c}\sin^2(\frac{\theta}{2})) \boldsymbol{E}_r(f,\theta_1,\theta_2) df  \,.
\end{equation}
Applying the matched filter on $\boldsymbol{E}_r(f,\theta_1,\theta_2)$, it only passes $\frac{2D}{c}\sin^2(\frac{\theta_i}{2})$ Fourier components and filters out the third term in \eqref{detailedspecs}.  Measuring $\boldsymbol{E}(\theta)$ for the whole range of possible DoAs ($[0^o,180^o]$), we can estimate DoAs and their powers by finding local maximum of $\boldsymbol{E}(\theta)$. 

As Fig. \ref{figtheta} illustrates, $\boldsymbol{E}(\theta)$ shows two distinct picks at $60^o$ and $100^o$ with $6$ dB difference between their amplitudes. Thus, $\boldsymbol{E}(\theta)$ is sufficient to estimate the DoAs and powers of the incoming paths.  
\begin{figure}[h!]
    \centering
    \includegraphics[width=4in,height=1.6in]{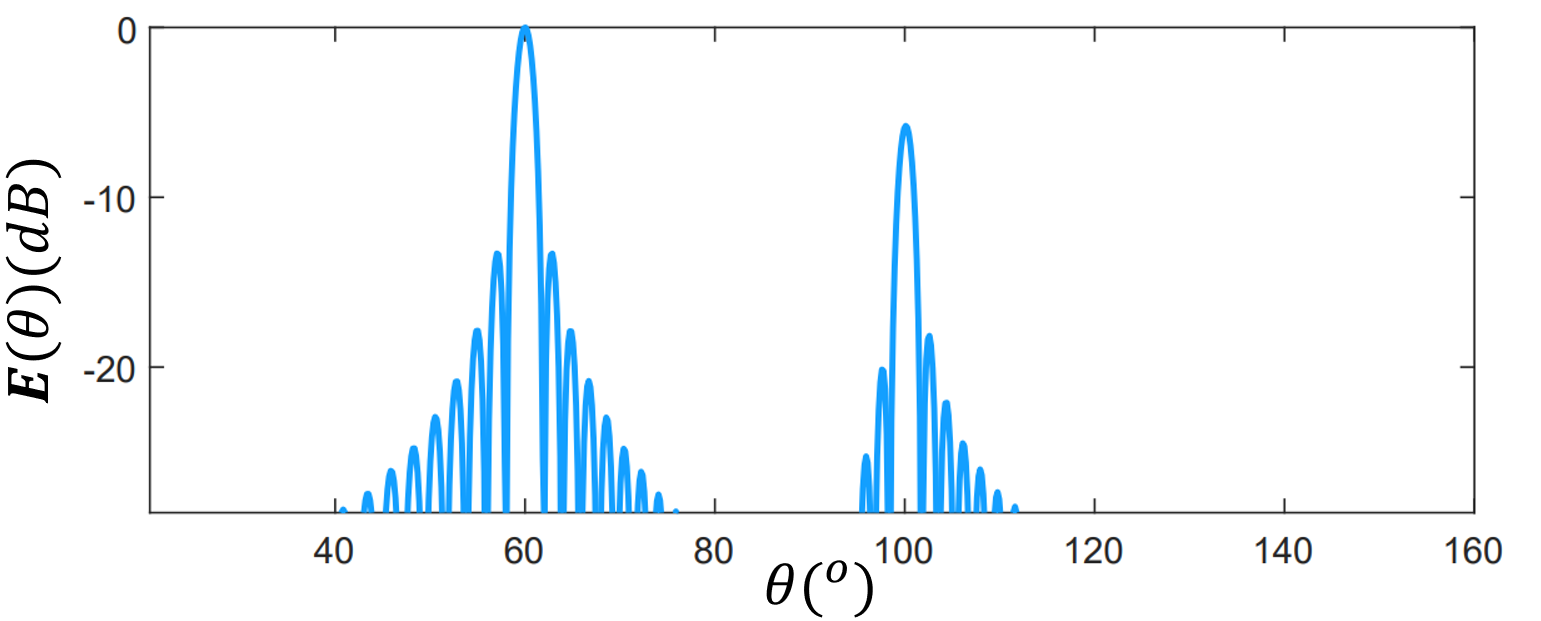}
    \caption{The output of applying the matched filter of expected received harmonic  response on $\boldsymbol{E}_r(f,\theta_1,\theta_2)$. {The $60^o$ and $100^o$ show the DoAs of two incoming signals to the antenna pair, from a LoS path and a NLoS path, respectively. }} 
    \label{figtheta}
\end{figure}

Next, to find the the relative ToF and relative distance between the two paths we can employ the spectrum of $\boldsymbol{E}_r(f,\theta_1,\theta_2)$. Fig. \ref{figspec} depicts $\boldsymbol{E}_r(f,\theta_1,\theta_2)$ and the absolute value of the Fourier transform of it. Referring to \eqref{detailedspecs}, the third term of $\boldsymbol{E}_r(f,\theta_1,\theta_2)$ shows a harmonic with $T^{\prime}_{1}-T^{\prime}_{2}$ frequency. Considering $\frac{2D}{c} \ll (T_{1}-T_{2}) $, we have $T^{\prime}_{1}-T^{\prime}_{2} \approx T_{1}-T_{2}$. Applying Fourier transform over $\boldsymbol{E}_r(f,\theta_1,\theta_2)$ and adjusting the frequency  by multiplying it by $c$, we can observe a pick at $c(T_1-T_2)$ which equals to the relative distance between the two paths. 
\begin{figure}[h!]
    \centering
    \includegraphics[width=4in,height=3.4in]{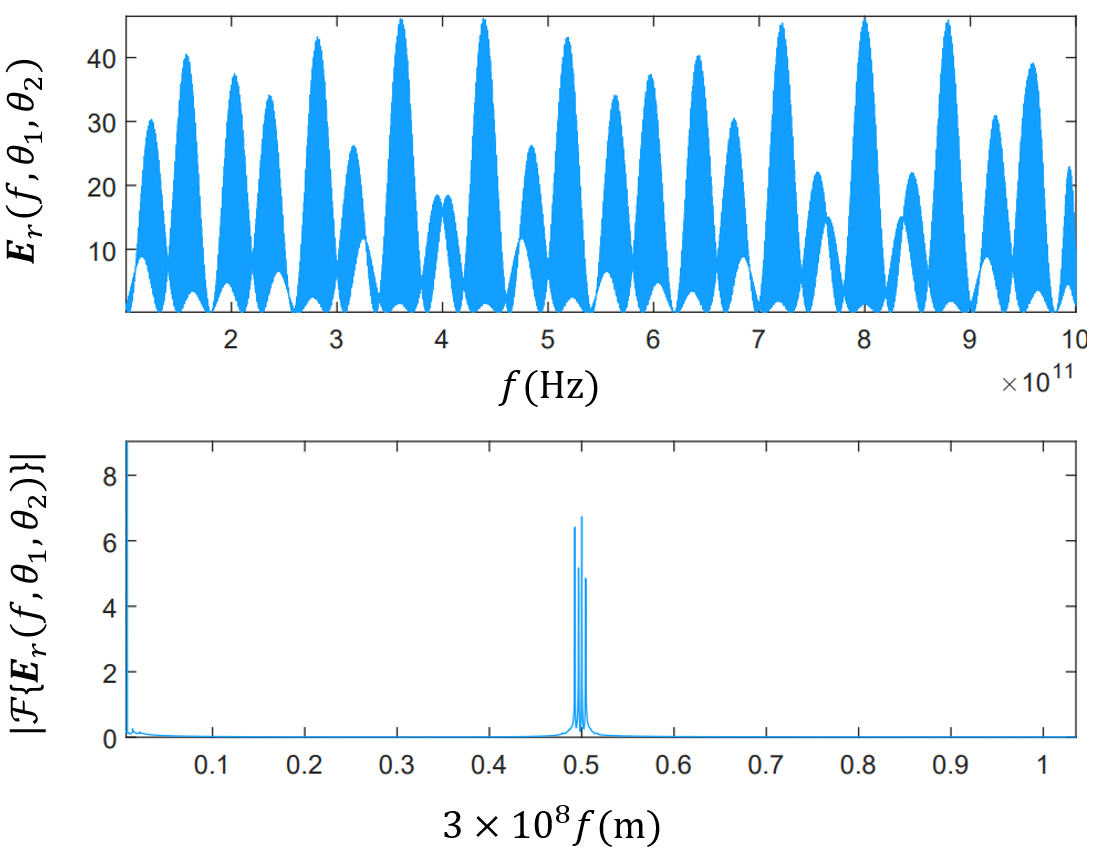}
    \caption{The spectrum and the spectrum of spectrum of the received signal at THz-TDS receiver { versus frequency (top) and frequency multiplied by light speed (bottom), respectively} assuming two paths are received at the receiver with 0.5m difference in path length.} 
    \label{figspec}
\end{figure}

As Fig. \ref{figspec} illustrates, a distinct pick at $0.5$ m is observable at the spectrum of $\boldsymbol{E}_r(f,\theta_1,\theta_2)$, which is equal to the relative distance between the two paths. Thus using the proposed technique, we could estimate DoAs, powers, and relative distance between the two paths.  By obtaining this information, and exploiting the Cost 2100 MIMO channel model \cite{liu2012cost}, we can estimate the channel state information (CSI) between the RX and the TX. 
Even in sub-6 GHz regime, estimating CSI in a single shot has not been possible and requires exchanging multiple pilot signals between the two side of the link \cite{van2017massive}. Therefore, not only SSH can be used to enable DoA estimation and beam selection for THz massive MIMO systems, but it also can assist the system by providing CSI information in a single shot. 

\subsection{Discussion}
As analytical developments and simulation results have shown, SSH is capable of providing CSI between a BS and the user. Referring to \cite{ali2017millimeter}, finding CSI (or the knowledge of the dominant propagation paths) is sufficient to find the best beam pair at TX and RX for the sake of maximizing communication rate. Thus, SSH can potentially eliminate  the need for exhaustive beam search for finding the optimal beam pair. Accordingly, SSH utterly quashes the burden of beam selection on the wireless communication and improves the system throughput radically. Further, SSH takes advantage of a very simple structure that is accessible via current fabrication technology, while fully digital or hybrid ULA and LA counterparts require complex and costly structures to conduct the same task. All in all, single-shot techniques and specifically SSH can revolutionize beam management at THz band and help us realize the full potential of ultra-fast communications feasible using THz wireless systems.      

\section{Conclusion}
\label{conc}
In this paper we have introduced a novel direction of arrival (DoA) estimation technique for THz wireless systems. We have shown that the proposed technique is capable of measuring line of sight (LoS) and multiple non-LoS DoAs simultaneously in a single-shot measurement. We have introduced a modification to the design that provides it with the capability of measuring the angle of departure (AoD) of the signal from the TX. We have shown by simulation that the proposed technique performs similar to a large ULA and LA in terms of angular resolution in the presence of noise. Further, we have demonstrated that we can increase the angular resolution of the technique by widening the gap between its two antennas without any additional hardware adjustment.     
}

\bibliographystyle{IEEEbib}
\bibliography{refs}
\end{document}